\documentclass[aps,showpacs,twocolumn,pra,superscriptaddress]{revtex4-2}

\usepackage[tbtags]{amsmath}
\usepackage{amssymb}
\usepackage{amsfonts}
\usepackage{graphicx}
\usepackage{bmpsize}
\usepackage{mathtools}
\usepackage[colorlinks=true,citecolor=blue,urlcolor=blue]{hyperref}
\usepackage{appendix}
\usepackage{times}
\hypersetup{colorlinks,linkcolor=blue,filecolor=blue,urlcolor=blue,citecolor=blue,}
\usepackage{multirow}

\begin{document}

\title{%Nonreciprocal Quantum Battery under Nonlinear Two-Photon Driving
Enhanced Nonreciprocal Quantum Battery Performance via Nonlinear Two-Photon Driving}

\author{Luxin Xu}
\affiliation{Key Laboratory of Low-Dimension Quantum Structures and Quantum Control of Ministry of Education, Synergetic Innovation Center for Quantum Effects and Applications, Xiangjiang-Laboratory, Institute of Interdisciplinary Studies and Department of Physics, Hunan Normal University, Changsha 410081, China}

\author{Changliang Ren}\thanks{Corresponding author: renchangliang@hunnu.edu.cn}
\affiliation{Key Laboratory of Low-Dimension Quantum Structures and Quantum Control of Ministry of Education, Synergetic Innovation Center for Quantum Effects and Applications, Xiangjiang-Laboratory, Institute of Interdisciplinary Studies and Department of Physics, Hunan Normal University, Changsha 410081, China} 
\affiliation{Hunan Research Center of the Basic Discipline for Quantum Effects and Quantum Technologies, Hunan Normal University, Changsha 410081, China}
\date{\today}

\begin{abstract}
%Quantum batteries, as highly efficient devices for energy storage and extraction, have garnered significant attention in recent years. This paper introduces a new quantum battery model based on nonreciprocal and nonlinear two-photon driving, which facilitates an efficient unidirectional charging mechanism through environmental engineering. By modeling the dynamics of the quantum battery using the Markov master equation, we derive an analytical solution for the system's time evolution and determine the parameter range at which dynamic equilibrium is achieved. The results indicate that increasing the strength of the driving field improves both energy conversion and storage efficiency, though it also extends the time required to achieve dynamic equilibrium. In comparison to single-photon driving, the two-photon system shows a clear advantage in storage capacity and entropy, with these advantages strengthening as the driving strength increases. Under asymmetric dissipation conditions, optimizing the system–environment coupling parameters can effectively improve quantum battery performance.  Furthermore, the proposed model is shown to be feasible for implementation across a variety of quantum experimental platforms, including photonic systems, superconducting circuits and magnonic platforms.
Quantum batteries have attracted significant attention as efficient quantum energy storage devices. In this work, we propose a nonlinear {\color{black}two-photon driving} quantum battery model featuring nonreciprocal dynamics that enables a highly efficient unidirectional charging mechanism through environmental engineering. Using a Markovian master-equation approach, we derive analytical solutions for the system dynamics and identify the parameter regime required for dynamical equilibration. Our results reveal that increasing the driving strength enhances both energy conversion and storage efficiency, albeit at the cost of longer equilibration times. Compared with single-photon driving, the two-photon process exhibits a pronounced advantage in energy capacity and entropy regulation, which becomes more prominent under stronger driving. Under asymmetric dissipation, optimizing the system--bath coupling can further improve performance. The proposed model is experimentally feasible and can be implemented across multiple quantum platforms, including photonic systems, superconducting circuits, and magnonic devices.
\end{abstract}
\maketitle
\section{Introduction}

In recent years, quantum thermodynamics has emerged as a key interdisciplinary field bridging microscopic quantum behavior with macroscopic thermodynamic laws \cite{PhysRevLett.115.210403, BRLW17, BENENTI20171}, driving rapid advances in quantum energy technologies \cite{mayer2023generalized,PhysRevLett.105.130401,rossnagel2016single,maslennikov2019quantum,PhysRevE.105.054115}. Within this context, quantum batteries (QBs)--quantum systems capable of storing and transferring energy through the manipulation of quantum states--have attracted growing interest \cite{PhysRevE.87.042123,RevModPhys.96.031001,PhysRevResearch.7.013151,hu2025enhancing,PhysRevLett.111.240401,PhysRevE.111.014121}. Unlike classical batteries, QBs exploit quantum resources such as coherence and entanglement to enhance charging power, energy utilization, and storage capacity \cite{PhysRevE.87.042123,PhysRevLett.120.117702,PhysRevLett.122.047702,PhysRevLett.131.240401,PhysRevLett.134.180401,PhysRevLett.129.130602,PhysRevA.106.062609,zhang2024entanglement}. While early studies focused primarily on two-level systems charged by classical fields, their intrinsic energy capacity remains fundamentally limited \cite{PhysRevE.99.052106,hu2022optimal,PhysRevResearch.6.023091,hu2022optimal,PhysRevE.99.052106}. To overcome this constraint, recent works have proposed multilevel and harmonic-oscillator-based QBs \cite{PhysRevB.98.205423,PhysRevResearch.2.033413,shaghaghi2022micromasers,gangwar2024coherently,PhysRevResearch.5.013155,PhysRevLett.127.100601}, which offer precise control over discrete energy spectra and can be realized across diverse experimental platforms, including photonic, superconducting, and magnonic systems.

As open quantum systems, QBs are inherently subject to environmental dissipation and decoherence, posing a central challenge to efficient energy storage and extraction \cite{kamin2023quantum,PhysRevB.98.205423,delmonte2021characterization,PhysRevA.102.052223}. Since complete isolation from the environment is practically impossible, recent research has focused on leveraging engineered environments to mitigate or even exploit dissipative effects \cite{vqnk-kzqg,PhysRevE.111.024125,PhysRevApplied.14.024092,PhysRevE.108.064106}. Approaches based on dark states \cite{PhysRevApplied.14.024092}, feedback control \cite{PhysRevE.104.044116, PhysRevE.106.014138}, Floquet engineering \cite{PhysRevA.102.060201}, and reservoir engineering  \cite{PhysRevLett.132.090401,fn1b-2m9g, PhysRevE.104.064143,PhysRevLett.132.210402,PhysRevApplied.23.024010} have demonstrated significant performance enhancement. For example, Quach et al. showed that dark-state coupling between noninteracting spins and their environment can substantially improve energy capacity and output power \cite{PhysRevApplied.14.024092}. Kamin et al. further analyzed strong system--bath coupling with fermionic reservoirs, revealing stable nonequilibrium energy accumulation \cite {PhysRevA.102.052223, Kamin_2020,PhysRevA.109.022226}. {\color{black} Energy backflow analysis of pseudomode-based models uncovers the impact of the environment on the charge-discharge dynamics of open QBs\cite{PhysRevE.111.024125}. Moreover, environment-engineered nonreciprocity mechanism enables directional energy transfer from the charger to the battery via shared dissipative and coherent channels, and has been investigated extensively\cite{PhysRevLett.132.210402,PhysRevApplied.23.024010,PhysRevX.5.021025}.}
%Moreover, pseudomode-based models have elucidated how environmental backflow and engineered nonreciprocity enable directional energy transfer from charger to battery via shared dissipative and coherent channels \cite{pusz1978passive,xu2023charging}. 
The innovation of this method \cite{PhysRevLett.132.210402,PhysRevApplied.23.024010} is to introduce nonreciprocal couplings through a shared environment and, by exploiting interference-like control, to precisely tune the shared dissipative and coherent pathways, thereby guiding the system into selected quantum states or evolution paths and enabling directed transport and storage of energy.

In addition, nonlinear two-photon driving has also emerged as a promising route to enhance QB performance by inducing quantum squeezing \cite{downing2023quantum,PhysRevA.109.052206}, dynamical instabilities\cite{leghtas2015confining}, and dissipative phase transitions\cite{PhysRevA.96.033826, PhysRevLett.122.110405}. This raises a key question: Can nonreciprocity effectively enhance the energy storage and charging efficiency of nonlinearly driven quantum batteries?

Motivated by these insights, we propose {\color{black}a nonlinear two-photon driving nonreciprocal quantum battery(QB)} model. A classical drive utilizes a second-order nonlinearity to inject photon pairs into the charging mode, with the battery dynamics governed by the Lindblad master equation under the Markov approximation. We derive analytical steady-state solutions below a critical driving threshold, beyond which the system loses stability, and obtain {\color{black}the precise} expressions for the stored energy, ergotropy, and output power. Our analysis reveals that nonreciprocity is crucial for enabling directed energy flow and significantly improving charging efficiency and energy extraction. This work underscores the fundamental role of nonreciprocal interactions in nonlinear quantum batteries and establishes a theoretical framework for advanced quantum energy-transfer technologies.

This paper is structured as follows. In Sec.~\ref {Sec2}, we establish a quantum battery model driven by a nonlinear two-photon. The dynamical evolution equations are derived, and based on these equations, analytical solutions for key physical quantities are obtained. In Sec.~\ref {Sec3}, the impact of different conditions on battery performance and the mechanisms for controlling these effects are analyzed. Finally, we conclude in Sec.~\ref {Sec4} with a summary.
\section{Theoretical Framework}\label{Sec2}
\subsection{The QB Model}
\begin{figure}[htbp]
	\centering
	\includegraphics[width=1\columnwidth]{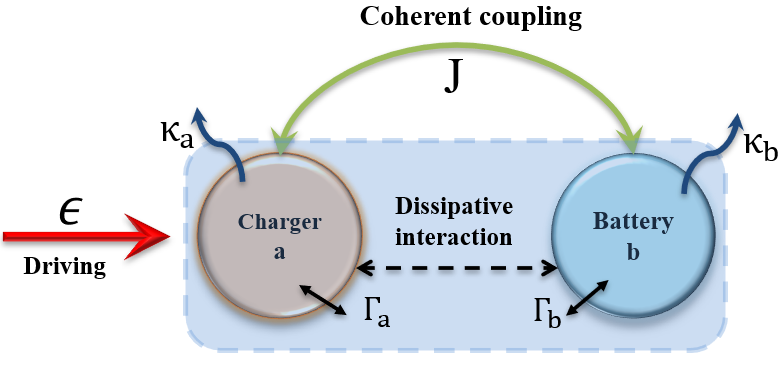}
	\caption{Illustration of {\color{black} two-photon driving quantum battery system}: The charger is subject to a parametric drive (indicated by the red arrow), characterized by amplitude $\epsilon$, phase $\theta$, and frequency $2\omega_p$. The charger $a$ and the battery $b$ are directly coupled via a coherent Hamiltonian $\hat{H}_{\text{coh}}$. In addition, both modes are connected to a common dissipative environment, with decay rates $\Gamma_a$ and $\Gamma_b$, respectively. By tuning the relative strengths of the coherent and dissipative interactions, one can effectively break the system's reciprocity. The local damping rates are denoted by $\kappa_a$ and $\kappa_b$, corresponding to the charger and the battery, respectively.}
	\label{fig:1}
\end{figure}

The model, schematically illustrated in Fig.~\ref{fig:1}, consists of two coupled quantum resonators: one functioning as a charger that absorbs external energy, and the other as a battery that stores it. Energy transfer between them occurs through a coherent coupling of strength $|J|$. A classical driving field with frequency $2\omega_p$, phase $\theta$, and amplitude $\epsilon$ continuously injects energy into the charger via a second-order nonlinear process. The charger and battery have natural frequencies $\omega_a$ and $\omega_b$, and local dissipation rates $\kappa_a$ and $\kappa_b$, respectively. The total Hamiltonian of the system can thus be decomposed into four distinct components:
%The model we study, illustrated schematically in Fig.\ref{fig:1}, comprises two coupled quantum resonators. Here, one resonator acts as a charger for energy input, and the other serves as a battery that stores energy. Their energy transfer is facilitated by a coherent coupling strength $|J|$. Furthermore, a classical driving field with frequency $2\omega_p$, phase $\theta$, and amplitude $\mathcal{\epsilon }$ is applied to continuously supply external energy to the charger. The natural frequencies of the charger and the battery are $\omega_a$ and $\omega_b$, respectively, with their corresponding local dissipation rates being $\kappa_a$ and $\kappa_b$. The total Hamiltonian can be systematically decomposed into four contributions: 
\begin{equation}\label{eq1}
	\mathrm{\hat{H} }= \mathrm{\hat{H}_{a}}+\mathrm{\hat{H}_{b}}+\mathrm{\hat{H}_{coh}}+\mathrm{\hat{H}_{d}},
\end{equation} 
where $\mathrm{\hat{H}_{a}} = \omega_{a} \hat{a}^{\dagger}\hat{a}$ and $\mathrm{\hat{H}_{b}} = \omega_{b} \hat{b}^{\dagger}\hat{b}$ describe the free Hamiltonians of the charger and battery, respectively. The term $\mathrm{\hat{H}_{coh}}=J\hat{a}^{\dagger}\hat{b}+J^{\ast  }\hat{a}\hat{b}^{\dagger}$ represents a coherent coupling between the two subsystems, where the complex parameter $J$ quantifies the coupling strength. Specially, the Hamiltonian of the quadratically driving field can be described as
\begin{equation}\label{eq5}
	\mathrm{\hat{H}_{d}}=\epsilon (\mathrm{e}^{i\theta}\mathrm{e}^{-2i\omega _{p}t}\hat{a}^{\dagger}\hat{a}^{\dagger}+\mathrm{e}^{-i\theta}\mathrm{e}^{2i\omega _{p}t}\hat{a}\hat{a}),
\end{equation}
where $\hat{a}^\dagger$ ($\hat{a}$) and $\hat{b}^\dagger$ ($\hat{b}$) are the creation (annihilation) operators for the charger and battery, respectively. 
We assume both resonators share the same characteristic frequency, $\omega_a = \omega_b = \omega$. The quadratic driving field is defined by its controllable amplitude $\epsilon$, phase $\theta$, and frequency $2\omega_p$.
%We assume that they have the same characteristic frequency $\omega_a = \omega_b = \omega$. The quadratically driving field, on the other hand, is characterized by an amplitude  $\epsilon$, a phase  $\theta$, and a frequency $2\omega_{p}$, all of which are controllable external parameters.
\par
For simplicity, we analyze the system in a rotating frame defined by the drive frequency $\omega_p$, where the effective Hamiltonian is given by:
\begin{equation}\label{eq6}
	\mathrm{\hat{H}_{I}}=\delta \hat{a}^{\dagger}\hat{a}+\delta \hat{b}^{\dagger}\hat{b}+J\hat{a}^{\dagger}\hat{b}+J^{\ast } \hat{a} \hat{b}^{\dagger}+\epsilon (\mathrm{e}^{i\theta}\hat{a}^{\dagger}\hat{a}^{\dagger}+\mathrm{e}^{-i\theta}\hat{a}\hat{a}),	
\end{equation} 
where $\delta = \omega - \omega_p$. The nonlinear Hamiltonian term involves a two-photon driving process, endowing the system with quantum squeezing characteristics.
%The nonlinear term in the above Hamiltonian incorporates a two-photon drive, which indicates the quantum-squeezing nature of the system.
\subsection{Dynamical Equations for QB}
%In this work, the charger and battery interact via two distinct coupling mechanisms. 
%To achieve nonreciprocal energy transport, we introduce a damped auxiliary cavity as a shared dissipative environment to which both the charger and the battery are coupled. By precisely balancing the dissipative interactions with their corresponding coherent couplings, unidirectional transport is realized. 
Nonreciprocal energy transport is achieved by introducing a damped auxiliary cavity that serves as a shared dissipative reservoir for both the charger and the battery. By finely tuning the balance between coherent and dissipative couplings, the system enables directional energy flow. Under the Markovian approximation\cite{PhysRevX.5.021025, PhysRevApplied.23.024010, PhysRevLett.132.210402,p93y-jflt, PhysRevB.99.035421}, the dynamics of the charger-battery subsystem is governed by a standard master equation for the reduced density matrix \(\tilde{\hat{\rho}}\), given as follows:
\begin{align}\label{eq7}
	\dot{\tilde{\hat{\rho}} } =-i \left [\mathrm{\hat{H}_{I}},\tilde{\hat{\rho}}  \right ] + \sum_{i=a,b} \kappa _{i}\mathcal{L}\left [\hat{i} \right ]\tilde{\hat{\rho}} +\Gamma \mathcal{L}\left [\hat{c}\right ]\tilde{\hat{\rho}} , 
\end{align}
where $\hat{c} = p_a \hat{a} + p_b \hat{b}$, $p_a$ and $p_b$ represent the relative coupling strengths of the charger and the battery to the dissipative environment. In Eq.~$\eqref{eq7}$, the first term describes the coherent interaction between the charger and the battery. The second term accounts for local damping of each mode into its own bath. The last term describes the interaction with the common dissipative environment, which occurs at rate $\Gamma$ and includes the induced charger-battery dissipation. $\mathcal{L}[\hat{o}] \tilde{\hat{\rho}} = \hat{o} \tilde{\hat{\rho}} \hat{o}^\dagger - \frac{1}{2} \left( \hat{o}^\dagger \hat{o} \tilde{\hat{\rho}} + \tilde{\hat{\rho}} \hat{o}^\dagger \hat{o} \right) $ are the Lindblad superoperators accounting for dissipation, $\hat{o}\in \left \{\hat{a} ,\hat{a}^\dagger,\hat{b},\hat{b}^\dagger \right \} $. For the first-order moments, we can directly obtain,
\begin{equation}\label{eq9}
	\begin{aligned}
		\langle \dot{\hat{a}} \rangle &= -\left( \frac{\Lambda}{2} + i\delta \right) \langle \hat{a} \rangle - i\left( J + i\mu \frac{\Gamma}{2} \right) \langle \hat{b} \rangle 
		- 2i e^{i\theta}\epsilon \langle \hat{a}^{\dagger } \rangle ,\\
		\langle \dot{\hat{b}} \rangle &= -\left( \frac{\Delta}{2} + i\delta \right) \langle \hat{b} \rangle 
		- i\left( J^{\ast} + i\mu^{\ast} \frac{\Gamma}{2} \right) \langle \hat{a} \rangle.
	\end{aligned}
\end{equation}
Similarly, for the second-order moments, we can obtain
\begin{widetext}
	\begin{equation}\label{eq10}
	\begin{aligned}
	&\dot{\langle \hat{a}^{\dagger }\hat{a}\rangle}\  =-\Lambda \langle \hat{a}^{\dagger }\hat{a}\rangle-2\mathrm{Im} \left \{2\epsilon\mathrm{e}^{-i\theta}\langle \hat{a}\hat{a}\rangle \right \} -2\mathrm{Re} \{ i(J+i\mu \frac{\Gamma }{2} )\langle \hat{a}^{\dagger  }\hat{b} \rangle \} ,\\
	&\dot{\langle \hat{a}\hat{a} \rangle }\  =-(\Lambda+2i\delta  ) \langle \hat{a}\hat{a}\rangle-4i\epsilon e^{i\theta } \langle \hat{a}^{\dagger }\hat{a} \rangle -2i\epsilon e^{i\theta }-i(2J+i\mu\Gamma)\langle \hat{a}\hat{b} \rangle ,\\
	&\dot{\langle \hat{a}^{\dagger }\hat{b}\rangle }\  =- ( \frac{\Lambda }{2}+ \frac{\Delta  }{2} )\langle \hat{a}^{\dagger } \hat{b} \rangle +i( J^{\ast } -i\mu ^{\ast }\frac{\Gamma }{2} ) \langle \hat{b}^{\dagger }\hat{b}\rangle -i ( J^{\ast } +i\mu ^{\ast } \frac{\Gamma }{2} ) \langle \hat{a}^{\dagger }\hat{a} \rangle+2i\epsilon e^{-i\theta } \langle \hat{a}\hat{b} \rangle  ,\\
	&\dot{\langle \hat{a}\hat{b} \rangle }\  =(-2i\delta-\frac{\Lambda }{2}-\frac{\Delta }{2} )\langle \hat{a}\hat{b}\rangle - 2i\epsilon e^{i\theta }\langle \hat{a}^{\dagger } \hat{b}\rangle -i(J^{\ast }+i\mu^{\ast } \frac{\Gamma }{2} )\langle \hat{a}\hat{a} \rangle-i(J +i\mu\frac{\Gamma }{2} )\langle \hat{b}\hat{b}\rangle ,\\ 
	&\dot{\langle \hat{b}^{\dagger} \hat{b}\rangle }\  =-\Delta \langle \hat{b}^{\dagger}\hat{b}\rangle+2\mathrm{Re} \{ i(J-i\mu \frac{\Gamma }{2} ) \langle \hat{a}^{\dagger } \hat{b} \rangle  \},\\
	&\dot{ \langle \hat{b}\hat{b} \rangle }\  =-(2i\delta +\Delta)  \langle \hat{b}\hat{b}\rangle -i(2J^{\ast }+i\mu^{\ast } \Gamma)\langle \hat{a}\hat{b} \rangle ,\\
	\end{aligned}
	\end{equation}
\end{widetext}
where $\mu = -p_b p_a^{\ast }$, and $\Gamma_i = \Gamma |p_i|^{2}$ for $i = a, b$. Both $p_a$ and $p_b$ are nonzero and satisfy the normalization condition $|p_b p_a^{\ast }| = 1$.
%The parameters $\Gamma_i$ describe local damping induced by the dissipative environment. We define $\Lambda = \Gamma_a + \kappa_a$ and $\Delta = \Gamma_b + \kappa_b$. To achieve nonreciprocity, we set $J=  -i\mu(\Gamma / 2)$. The initial conditions are specified as: 
The parameters $\Gamma_i$ represent local damping induced by the dissipative environment. 
For convenience, we define the total decay rates as $\Lambda = \Gamma_a + \kappa_a$ and $\Delta = \Gamma_b + \kappa_b$. 
Nonreciprocity is introduced by setting the coherent coupling to $J = -i\mu(\Gamma/2)$. 
The initial conditions are given by $\langle \hat{a}\rangle\mid_{t=0}=\langle \hat{a}^{\dagger } \hat{a} \rangle\mid_{t=0}=\langle \hat{a}\hat{a}\rangle\mid_{t=0}=\langle \hat{a}^{\dagger}\hat{b}\rangle\mid_{t=0}=\langle \hat{a}\hat{b}\rangle\mid_{t=0}=0$ and $\langle \hat{b}\rangle\mid_{t=0}=\langle \hat{b}^{\dagger } \hat{b}\rangle\mid_{t=0}=\langle \hat{b}\hat{b}\rangle\mid_{t=0}=0$. Therefore, by solving the coupled differential equations Eq~$\eqref{eq9}$ and $\eqref{eq10}$ for the first- and second-order moments, we can fully capture the dynamics of the composite system $\tilde{\hat{\rho}}(t)$. For simplicity, we set $\delta = 0$, i.e.,
\begin{widetext}
	\begin{subequations}\label{eq12}
		\begin{align}
\left\langle b \right\rangle\big|_{t} &= 0, \\
\left\langle a^{\dagger} a \right\rangle\big|_{t} &= 
\frac{\mathrm{e}^{4 t \epsilon - t \Lambda}\,\epsilon}{4\epsilon - \Lambda}
+ \frac{8\,\epsilon^{2}}{\Lambda^{2} - 16\,\epsilon^{2}}
+ \frac{\mathrm{e}^{-t (4 \epsilon + \Lambda)}\,\epsilon}{4\epsilon + \Lambda}, \\
\left\langle b^{\dagger} b \right\rangle\big|_{t} &= 
\frac{64 \mathrm{e}^{-t \Delta} \Gamma^2 \epsilon^2 (\Delta - \Lambda)}{\Delta \left[ (\Delta - \Lambda)^2 - 16 \epsilon^2 \right]^2} 
+ \frac{4 \mathrm{e}^{4 t \epsilon - t \Lambda} \Gamma^2 \epsilon}{(4 \epsilon - \Lambda)(\Delta + 4 \epsilon - \Lambda)^2}+ \frac{16 \mathrm{e}^{-\frac{1}{2} t (\Delta - 4 \epsilon + \Lambda)} \Gamma^2 \epsilon}{(\Delta + 4 \epsilon - \Lambda)^2 (\Delta - 4 \epsilon + \Lambda)} 
+ \frac{4 \mathrm{e}^{-t (4 \epsilon + \Lambda)} \Gamma^2 \epsilon}{(4 \epsilon + \Lambda)(4 \epsilon + \Lambda - \Delta)^2}\notag \\
&\quad - \frac{16 \mathrm{e}^{-\frac{1}{2} t (\Delta + 4 \epsilon + \Lambda)} \Gamma^2 \epsilon}{(4 \epsilon + \Lambda - \Delta)^2 (\Delta + 4 \epsilon + \Lambda)} - \frac{32 \Gamma^2 \epsilon^2 (\Delta + 2 \Lambda)}{\Delta (16 \epsilon^2 - \Lambda^2) \left[ (\Delta + \Lambda)^2 - 16 \epsilon^2 \right]}, \\
\left\langle b b \right\rangle\big|_{t} &=  
\frac{8 i \mathrm{e}^{-i\theta} \Gamma^2 \epsilon \left(16 \epsilon^2 + \Lambda (\Delta + \Lambda) \right)}{\Delta \left[ (\Delta + \Lambda)^2 - 16 \epsilon^2 \right] (\Lambda^2 - 16 \epsilon^2)} 
- \frac{8 i \mathrm{e}^{-t \Delta - i \theta} \Gamma^2 \epsilon \left( 16 \epsilon^2 + (\Delta - \Lambda)^2 \right)}{\Delta \left[ (\Delta - \Lambda)^2 - 16 \epsilon^2 \right]^2} 
- \frac{4 i \mathrm{e}^{4 t \epsilon - i \theta - t \Lambda} \Gamma^2 \epsilon}{(\Delta + 4 \epsilon - \Lambda)^2 (\Delta + 4 \epsilon)} \notag \\
&\quad + \frac{16 i \mathrm{e}^{-i \theta - \tfrac{1}{2} t (\Delta - 4 \epsilon + \Lambda)} \Gamma^2 \epsilon}{(\Delta + 4 \epsilon - \Lambda)^2 (\Delta - 4 \epsilon + \Lambda)}
- \frac{4 i \mathrm{e}^{-i \theta - t (4 \epsilon + \Lambda)} \Gamma^2 \epsilon}{(4 \epsilon + \Lambda - \Delta)^2 (4 \epsilon + \Lambda)} 
+ \frac{16 i \mathrm{e}^{-i \theta - \tfrac{1}{2} t (\Delta + 4 \epsilon + \Lambda)} \Gamma^2 \epsilon}{(4 \epsilon + \Lambda - \Delta)^2 (\Delta + 4 \epsilon + \Lambda)}.
\end{align}
\end{subequations}
\end{widetext}
%From the expression of the analytical solution above, we can directly deduce the condition for the steady-state solution to occur, which is  $\epsilon< \frac{\Lambda }{4}$.
The analytical solution indicates that a stable steady state exists only when the driving strength fulfills  
\begin{equation}
    \epsilon < \frac{\Lambda}{4}.
\end{equation}\par
\subsection{Relevant Performance Indicators of QB}
The assessment of QB performance relies on two key physical quantities: \( E_{b} \) and \( E_{b}^{\beta} \). While \( E_{b} \) represents the total stored energy, not all of it can be extracted as useful work due to non-classical effects like quantum coherence. To quantify the non-extractable portion, \( E_{b}^{\beta} \) is introduced, defined as the energy that is irretrievable under any cyclic unitary operation. Collectively, \( E_{b} \) and \( E_{b}^{\beta} \) fully characterize the energy properties of QBs, as specified in the following expressions,
\begin{equation}\label{eq13}
	E_{b}=\mathrm{Tr}\left \{ \hat{H}_{b} \rho_{b}  \right \} =\omega_{b} \left \langle b^{\dagger }b \right \rangle,
\end{equation}
\begin{equation}\label{eq14}
	E_{b}^{\beta }=\mathrm{Tr}\left \{ \hat{H}_{b} \rho_{b}^{\beta }  \right \}=\omega_{b} \left ( \frac{\sqrt{\mathcal{D}}-1 }{2}  \right ), 
\end{equation}
where $\mathcal{D}=\left\{ 1 + 2\left\langle b^\dagger b \right\rangle - 2\langle b^\dagger\rangle \langle b\rangle \right\}^2 - 4 \left| \langle bb \rangle - \langle b \rangle^2 \right|^2$ and $\omega_b = \omega$. To quantify the energy storage performance of the battery, the ergotropy $\varepsilon_{b}$ can be defined as,
\begin{equation}\label{eq15}
	\varepsilon_{b} =E_{b}-E_{b}^{\beta }.
\end{equation}
Based on Eq.~$\eqref{eq12}$, the corresponding formulation is readily derived,
\begin{widetext}\label{eq16}
	\begin{subequations}
		\begin{align}
			E_{b} \big|_{t}  &=  
			\frac{64\omega \mathrm{e}^{-t \Delta} \Gamma^2 \epsilon^2 (\Delta - \Lambda)}{\Delta \left[ (\Delta - \Lambda)^2 - 16 \epsilon^2 \right]^2} 
			+ \frac{4\omega \mathrm{e}^{4 t \epsilon - t \Lambda} \Gamma^2 \epsilon}{(4 \epsilon - \Lambda)(\Delta + 4 \epsilon - \Lambda)^2}+ \frac{16 \omega \mathrm{e}^{-\frac{1}{2} t (\Delta - 4 \epsilon + \Lambda)} \Gamma^2 \epsilon}{(\Delta + 4 \epsilon - \Lambda)^2 (\Delta - 4 \epsilon + \Lambda)} 
			+ \frac{4\omega \mathrm{e}^{-t (4 \epsilon + \Lambda)} \Gamma^2 \epsilon}{(4 \epsilon + \Lambda)(4 \epsilon + \Lambda - \Delta)^2}\notag \\
			&\quad - \frac{16\omega \mathrm{e}^{-\frac{1}{2} t (\Delta + 4 \epsilon + \Lambda)} \Gamma^2 \epsilon}{(4 \epsilon + \Lambda - \Delta)^2 (\Delta + 4 \epsilon + \Lambda)} - \frac{32\omega  \Gamma^2 \epsilon^2 (\Delta + 2 \Lambda)}{\Delta (16 \epsilon^2 - \Lambda^2) \left[ (\Delta + \Lambda)^2 - 16 \epsilon^2 \right]},\\
		\varepsilon_{b}\big|_{t} &=  
			\frac{64\omega \mathrm{e}^{-t \Delta} \Gamma^2 \epsilon^2 (\Delta - \Lambda)}{\Delta \left[ (\Delta - \Lambda)^2 - 16 \epsilon^2 \right]^2} 
			+ \frac{4\omega \mathrm{e}^{4 t \epsilon - t \Lambda} \Gamma^2 \epsilon}{(4 \epsilon - \Lambda)(\Delta + 4 \epsilon - \Lambda)^2}+ \frac{16 \omega \mathrm{e}^{-\frac{1}{2} t (\Delta - 4 \epsilon + \Lambda)} \Gamma^2 \epsilon}{(\Delta + 4 \epsilon - \Lambda)^2 (\Delta - 4 \epsilon + \Lambda)} 
			+ \frac{4\omega \mathrm{e}^{-t (4 \epsilon + \Lambda)} \Gamma^2 \epsilon}{(4 \epsilon + \Lambda)(4 \epsilon + \Lambda - \Delta)^2}\notag \\
			&\quad - \frac{16\omega \mathrm{e}^{-\frac{1}{2} t (\Delta + 4 \epsilon + \Lambda)} \Gamma^2 \epsilon}{(4 \epsilon + \Lambda - \Delta)^2 (\Delta + 4 \epsilon + \Lambda)} - \frac{32\omega  \Gamma^2 \epsilon^2 (\Delta + 2 \Lambda)}{\Delta (16 \epsilon^2 - \Lambda^2) \left[ (\Delta + \Lambda)^2 - 16 \epsilon^2 \right]}-\omega \left ( \frac{\sqrt{x(t)y(t)}-1 }{2}  \right ),
		\end{align}
	\end{subequations}
where
	\begin{subequations}\label{eq17}
		\begin{align}
		x(t) &=  
			 1 
			- \frac{16 e^{-t \Delta} \Gamma^2 \epsilon}{\Delta (\Delta + 4\epsilon - \Lambda)^2}
			+ \frac{16 e^{4t\epsilon - t\Delta} \Gamma^2\epsilon}{\Delta (4\epsilon - \Lambda)(\Delta + 4\epsilon - \Lambda)^2}
			- \frac{16 \Gamma^2 \epsilon}{\Delta (4\epsilon - \Lambda)(\Delta - 4\epsilon+ \Lambda)}
			+ \frac{64 e^{-\frac{1}{2} t (\Delta - 4\epsilon + \Lambda)} \Gamma^2 \epsilon}{(\Delta + 4\epsilon - \Lambda)^2 (\Delta - 4\epsilon + \Lambda)},\\
			y(t) &=  
			1 
			+ \frac{16 e^{-t \Delta} \Gamma^2 \epsilon}{\Delta (4\epsilon + \Lambda - \Delta)^2}
			- \frac{16 \Gamma^2\epsilon}{\Delta (4\epsilon + \Lambda)(4\epsilon + \Lambda - \Delta)}
			+ \frac{16 e^{-t (4\epsilon + \Lambda)} \Gamma^2 \epsilon}{(4\epsilon+ \Lambda)(4\epsilon + \Lambda - \Delta)^2}
			- \frac{64 e^{-\frac{1}{2} t (\Delta + 4\epsilon + \Lambda)} \Gamma^2 \epsilon}{(4\epsilon + \Lambda - \Delta)^2 (\Delta + 4\epsilon + \Lambda)}.
		\end{align}
	\end{subequations}
\end{widetext}
 Furthermore, the instantaneous charging power is defined as the time derivative of the battery energy, $P_{\varepsilon_b}(t) = \frac{d\varepsilon_b(t)}{dt}$
. This quantity quantifies the rate of energy exchange, $P_{\varepsilon_b}(t) > 0$ corresponds to charging, whereas $P_{\varepsilon_b}(t) < 0$ denotes discharging.

 %Furthermore, we introduce the instantaneous charging power \(P_{\varepsilon_b}(t)\), which is defined as the time derivative of the battery energy, \(P_{\varepsilon_b}(t) = d\varepsilon_{b}(t)/dt\).This quantity characterizes the charge-discharge dynamics: \(P_{\varepsilon_b}(t)>0\) indicates charging, whereas \(P_{\varepsilon_b}(t)<0\) indicates discharging.
\section{Nonlinear Driving Effects on the Performance of Nonreciprocal QB under Symmetric and Asymmetric Dissipation}\label{Sec3}
\subsection{Performance of Nonreciprocal QB with Nonlinear Two-Photon Driving under Symmetric Dissipation}\label{Sec3A}\par
\begin{figure*}[htbp]
	\centering
	\includegraphics[width=0.685\columnwidth]{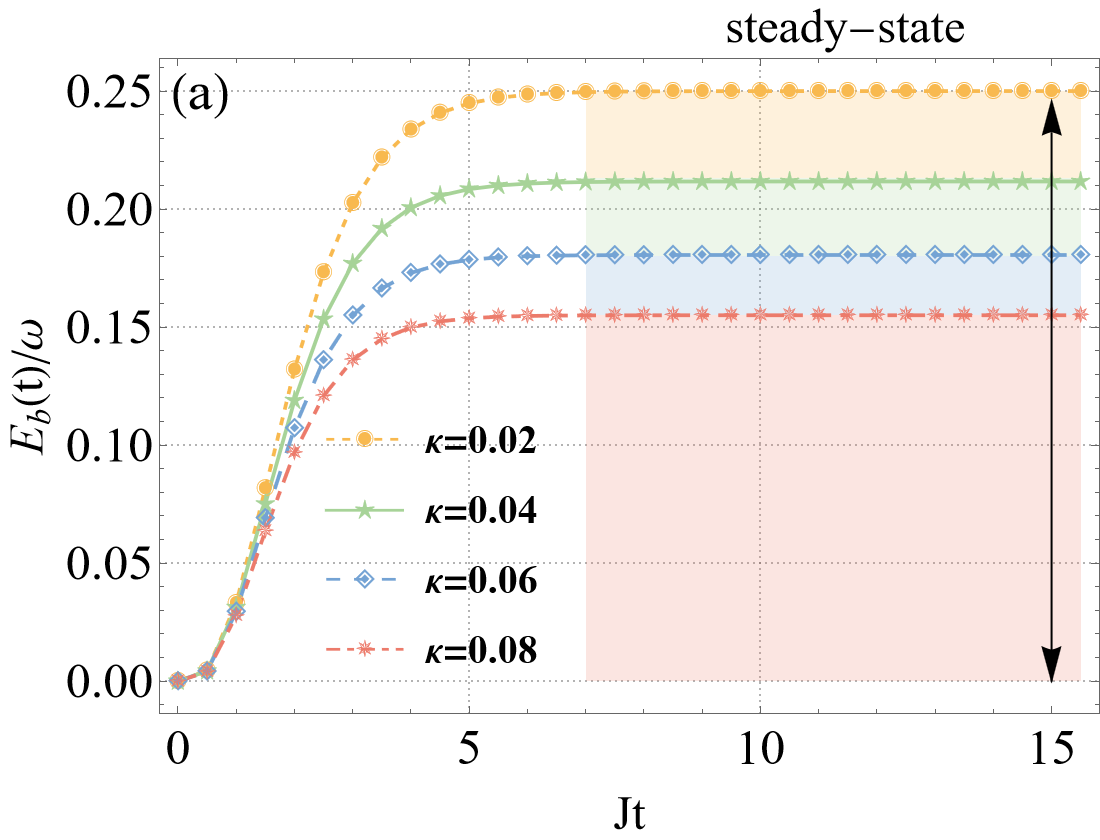}
	\includegraphics[width=0.685\columnwidth]{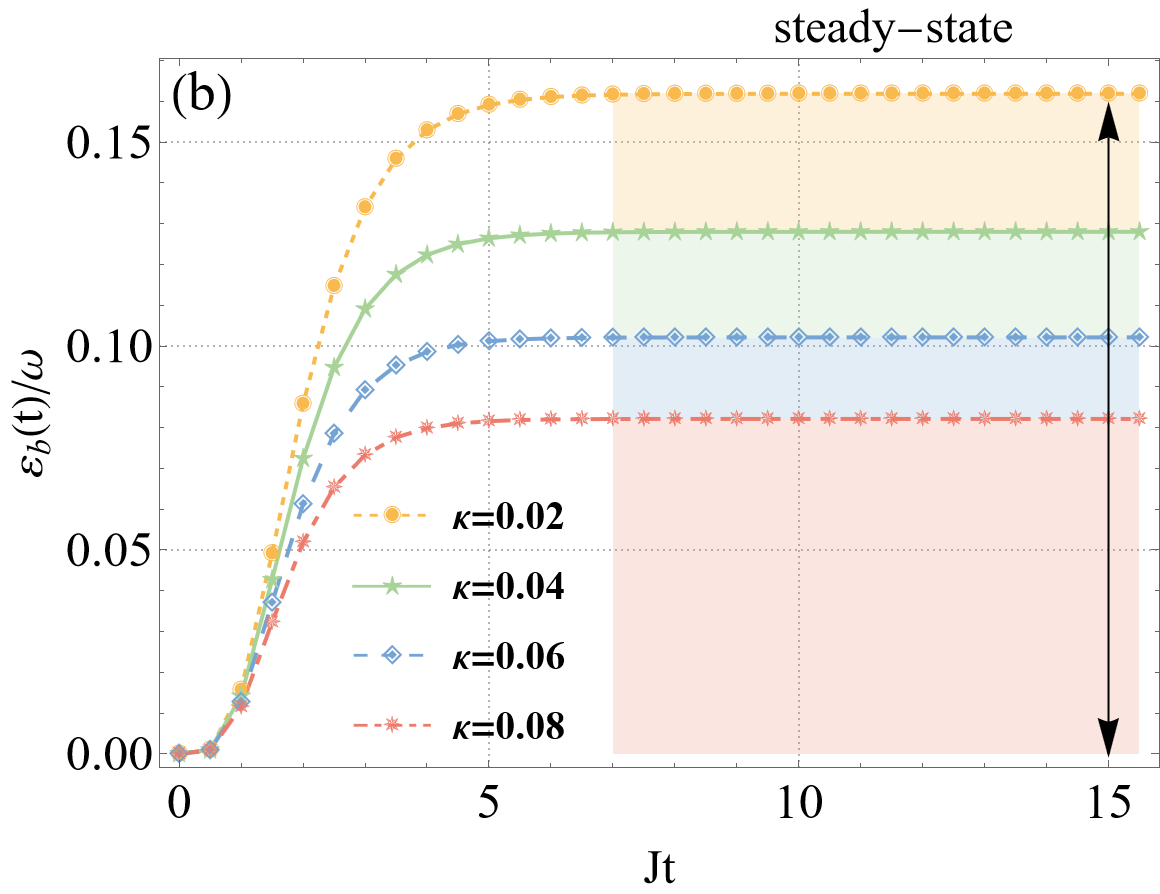}
	\includegraphics[width=0.67\columnwidth]{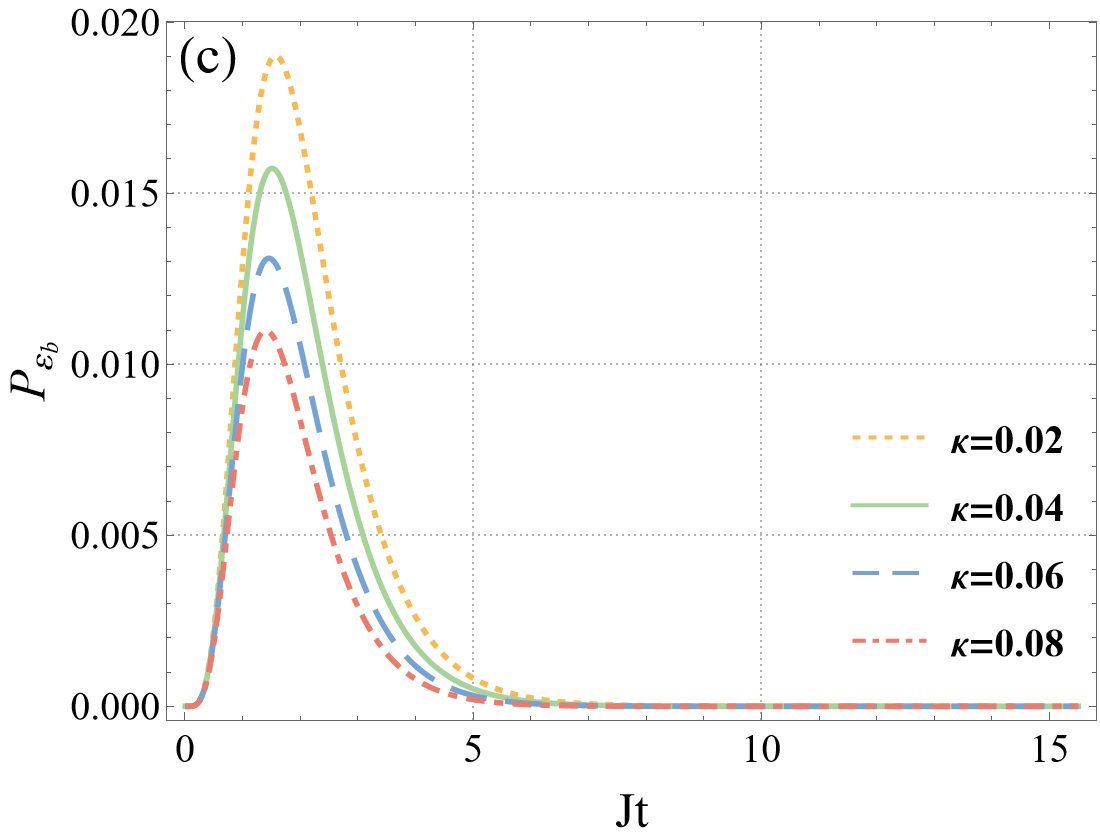}
	\includegraphics[width=0.67\columnwidth]{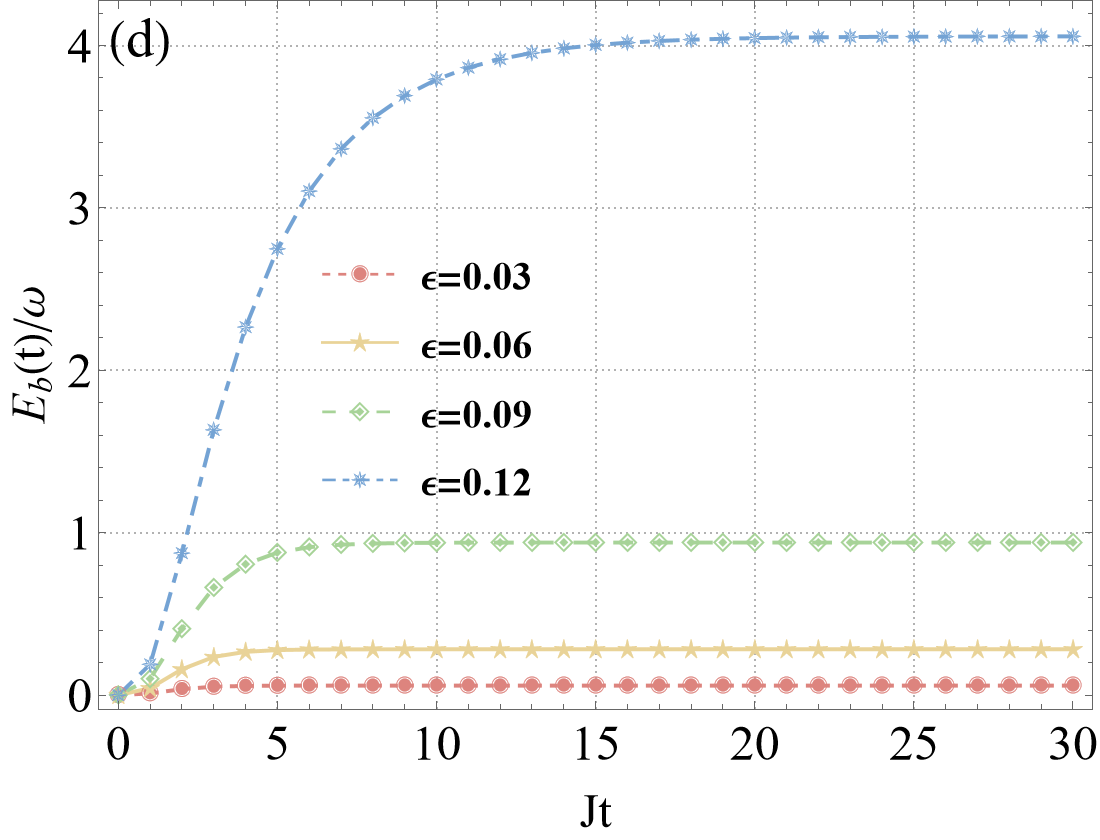}
	\includegraphics[width=0.67\columnwidth]{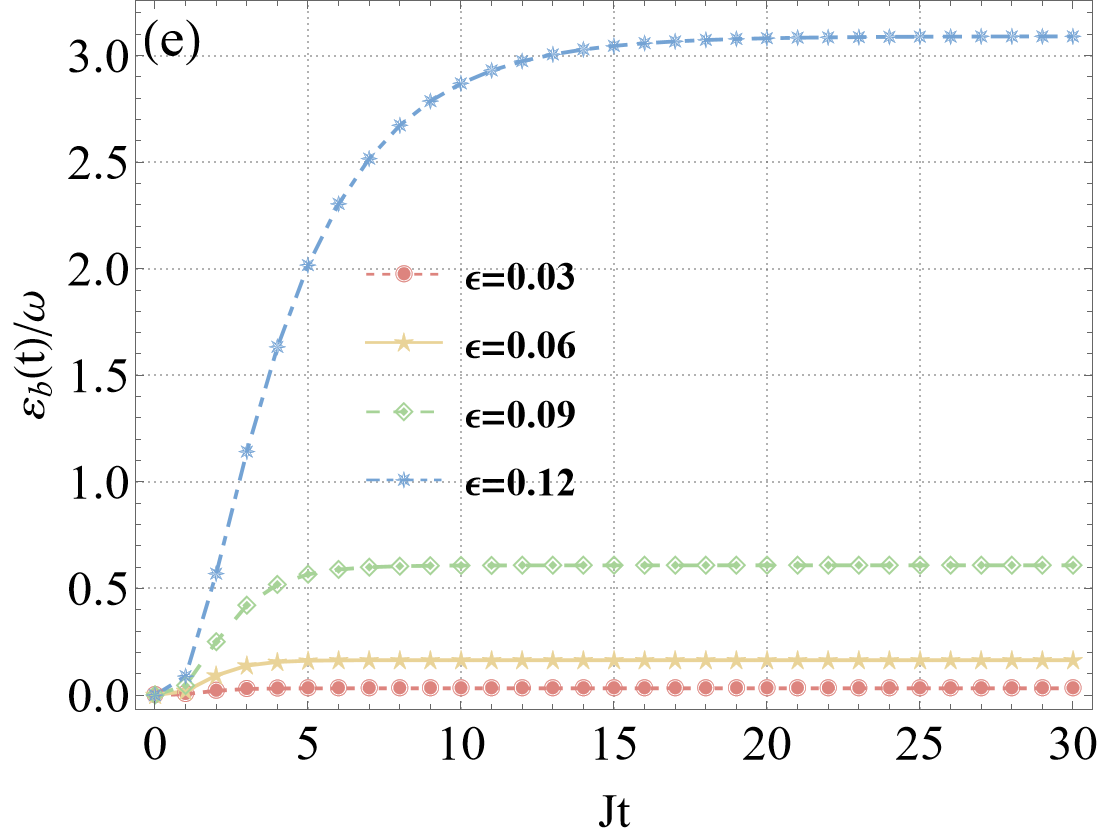}
	\includegraphics[width=0.67\columnwidth]{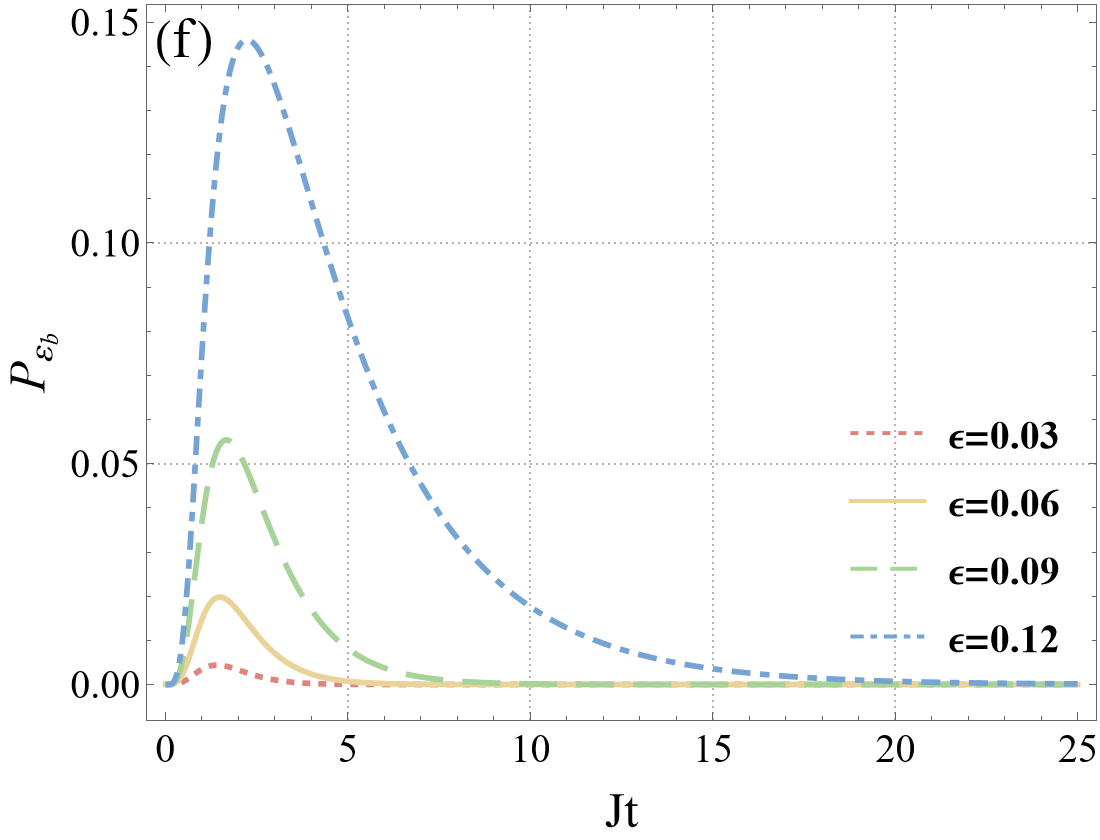}
	\caption{Figs (a)-(c) illustrate the dynamics of the QB's stored energy $E_b(t)/\omega$, ergotropy $\varepsilon(t)/\omega$, and the instantaneous power of the ergotropy $P_{\varepsilon b}$ as functions of the scaled time $Jt$, under varying local damping strengths $\kappa$. In this case, the driving amplitude is fixed at $\epsilon = 0.05\omega$. Figs (d)-(f) further explore how these three quantities evolve when the damping is fixed at $\kappa = 0.06\omega$ and the driving strength varies as $\epsilon/\omega = 0.03,\ 0.06,\ 0.09,\ 0.12$. In all panels, the horizontal axis represents the scaled time $Jt$.
		Other parameters are set as: $\kappa = \kappa_a = \kappa_b$, $\Gamma = \Gamma_a = \Gamma_b$, and $|J| = \Gamma/2 = 0.25\omega$.}
	\label{fig2}
\end{figure*}\par
Using environmental engineering, we construct a nonreciprocal QB system capable of unidirectional energy transport, 
%thus circumventing the limitations imposed by reciprocity in traditional architectures. To assess the charging performance of a nonlinear two-photon-driven nonreciprocal QB, we perform a systematic analysis of the energy-evolution dynamics during charging under symmetric conditions($\Lambda = \Delta$). 
thereby overcoming the reciprocity constraints inherent in conventional architectures. To evaluate the charging performance of a nonlinear two-photon {\color{black}driving} nonreciprocal QB, we analyze the energy-evolution dynamics under symmetric conditions ($\Lambda = \Delta$). 
%We first quantitatively analyze the influence of the local damping dissipation rate $\kappa$ on the energy in the QB.
Specifically, we examine the effect of the local damping rate $\kappa$ on the battery energy.
Figs.~\ref{fig2}(a) and~\ref{fig2}(b) illustrate the time evolution of the stored energy $E_b(t)/\omega$ and the ergotropy $\varepsilon_b(t)/\omega$ for various $\kappa$.
%Figs.\ref{fig2}~(a) and (b) depict the evolution of the stored energy $E_b(t)/\omega$ and the ergotropy $\varepsilon_{b}(t)/\omega$, respectively, as they evolve towards their steady states for different $\kappa$ values. 
The results reveal that both quantities are significantly suppressed as $\kappa$ increases, exhibiting a gradual reduction in their steady-state values. 
Both quantities decrease markedly with increasing $\kappa$, exhibiting lower steady-state plateaus as dissipation strengthens. 
This trend indicates that enhanced local damping facilitates energy leakage to the environment, thereby accelerating energy loss. 
Remarkably, the characteristic relaxation timescale $Jt$ at which dynamical equilibrium is achieved remains nearly independent of $\kappa$, suggesting that local damping primarily degrades steady-state performance while leaving the transient relaxation dynamics largely unaffected.
%This indicates that enhanced dissipative coupling promotes energy exchange between the system and its environment, consequently accelerating energy dissipation into the environment. Notably, the characteristic relaxation timescale $Jt$ at which dynamical balance is reached is essentially insensitive to $\kappa$, implying that local damping primarily degrades steady-state performance while leaving the relaxation dynamics largely unaffected. 
In the steady-state limit, the energy storage and ergotropy of QB are
\begin{subequations}\label{eq17}
		\begin{align}
		&E_{b}(\infty)/\omega =  
			 \frac{24\Gamma ^{2}\epsilon^{2}}{64\epsilon ^{4}-20\epsilon ^{2}\Lambda^{2}+\Lambda^{4}},\\
			&\varepsilon_{b}(\infty)/\omega =  
			\frac{24\Gamma ^{2}\epsilon^{2}}{64\epsilon ^{4}-20\epsilon ^{2}\Lambda^{2}+\Lambda^{4}}+\frac{1}{2}\notag \\
            &-\frac{1}{2} \left ( \sqrt{\frac{32 \epsilon^{2} \left( 3 \Gamma^{2} + 2 \epsilon^{2} \right) \Lambda^{2} -64 \Gamma^{4} \epsilon^{2}- 20 \epsilon^{2} \Lambda^{4}+ \Lambda^{6}}{64\epsilon ^{4}-20\epsilon ^{2}\Lambda^{2}+\Lambda^{4}}}\right) .
		\end{align}
	\end{subequations}
Fig.~\ref{fig2}(c) shows the evolution of the instantaneous charging power $P_{\varepsilon_b}$ for QB versus the characteristic time $Jt$. The instantaneous power $P_{\varepsilon_b}$ rises rapidly to a maximum value and subsequently decreases. At $Jt=8$, $P_{\varepsilon_b}=0$, marking the system's arrival at a dynamic equilibrium. We also investigate the impact of the external driving fields on the charging performance. For a fixed dissipation rate $\kappa=0.06\omega$, Figs.~\ref{fig2}(d) and~\ref{fig2}(e) display the evolutions of the stored energy and ergotropy with the driving strength $\epsilon$. The results demonstrate a marked enhancement in both stored energy and ergotropy with increasing $\epsilon$, underscoring the role of the enhanced drive in promoting external energy injection. Notably, the characteristic time scale $Jt$ for the system to attain dynamic equilibrium is prolonged with increasing $\epsilon$, finally approaching a constant value. As shown in Fig.~\ref{fig2}(f),  $P_{\varepsilon_b}$ rises rapidly to a peak in the initial stage and subsequently decays slowly to zero. With increasing $\epsilon$, the time required for $P_{\varepsilon_b}$ to reach its peak, as well as for the system to converge to dynamic equilibrium (characterized by $Jt$), is notably prolonged. This phenomenon suggests that a higher $\epsilon$ causes the energy input rate to surpass the system's internal relaxation. Consequently, QB must establish a new dynamic equilibrium through more complex internal dynamics, such as many-body interactions or nonlinear responses, which extends the characteristic timescale $Jt$. In a nonreciprocal QB, an enhanced driving field not only strengthens system coupling and quantum squeezing but also facilitates efficient energy storage and ergotropy.\par
\begin{figure*}[htbp]
	\centering
	\includegraphics[width=0.85\columnwidth]{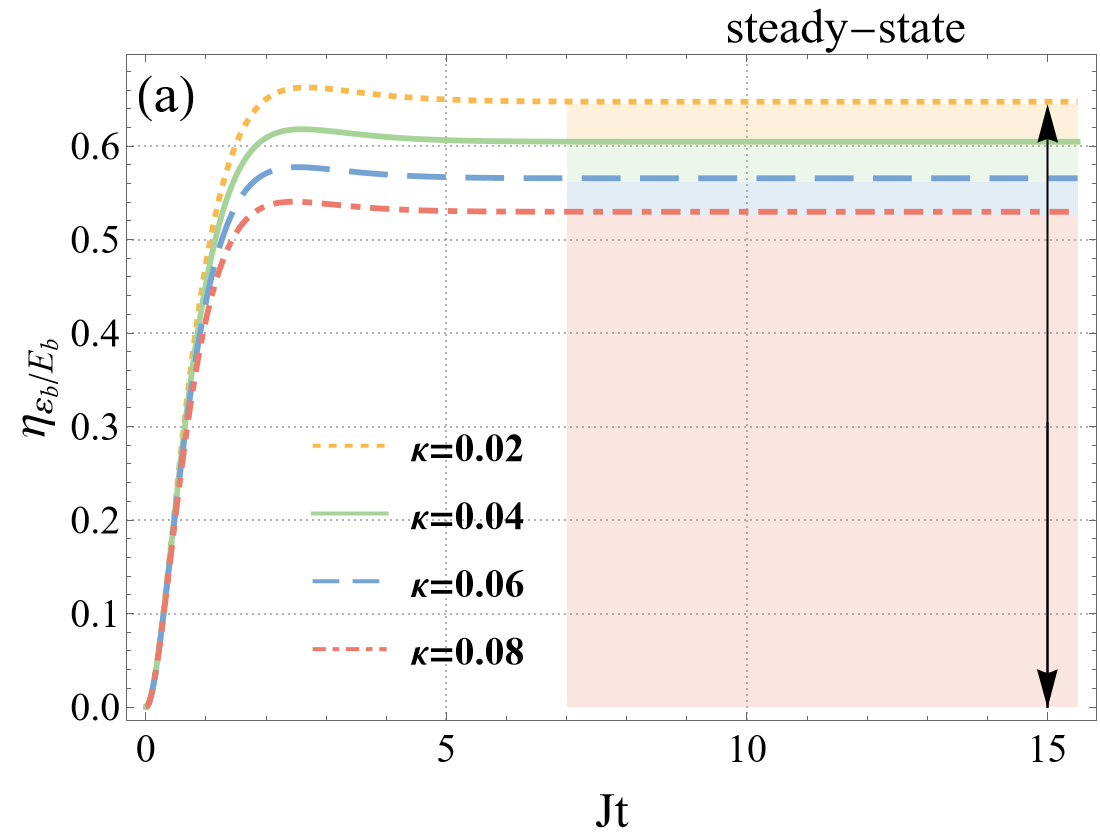}
	\includegraphics[width=0.825\columnwidth]{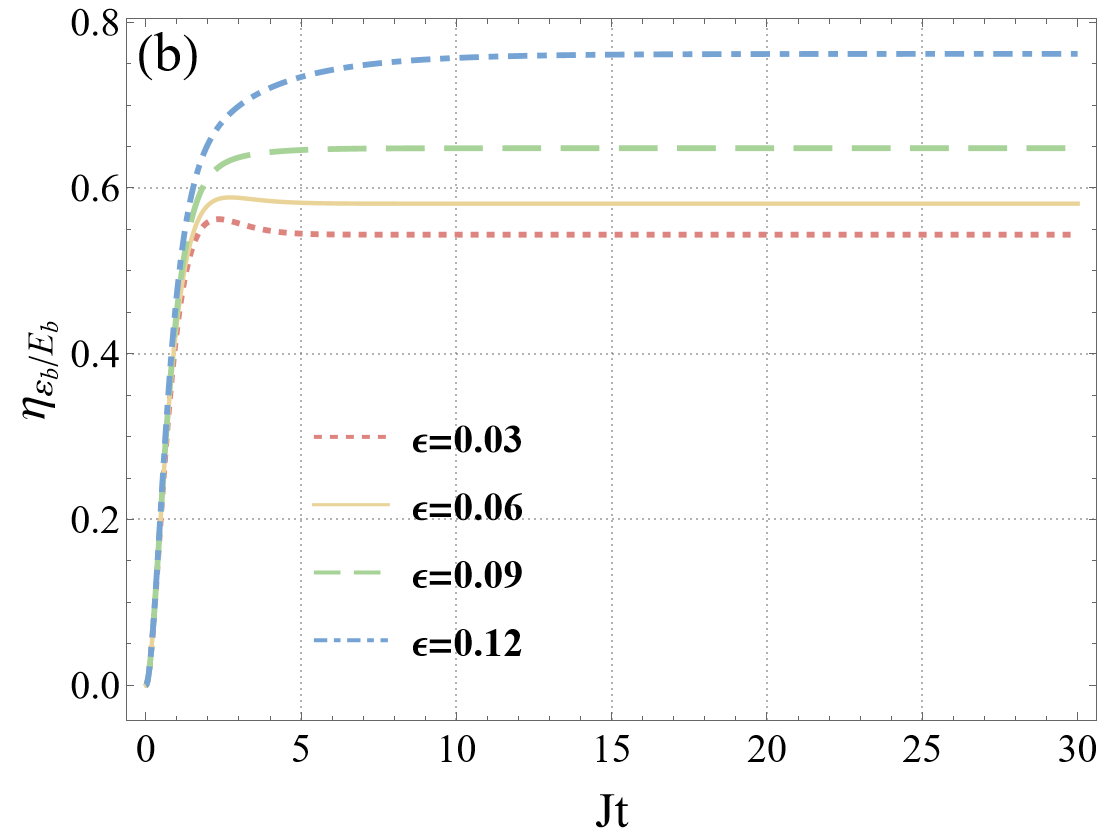}
	\includegraphics[width=0.85\columnwidth]{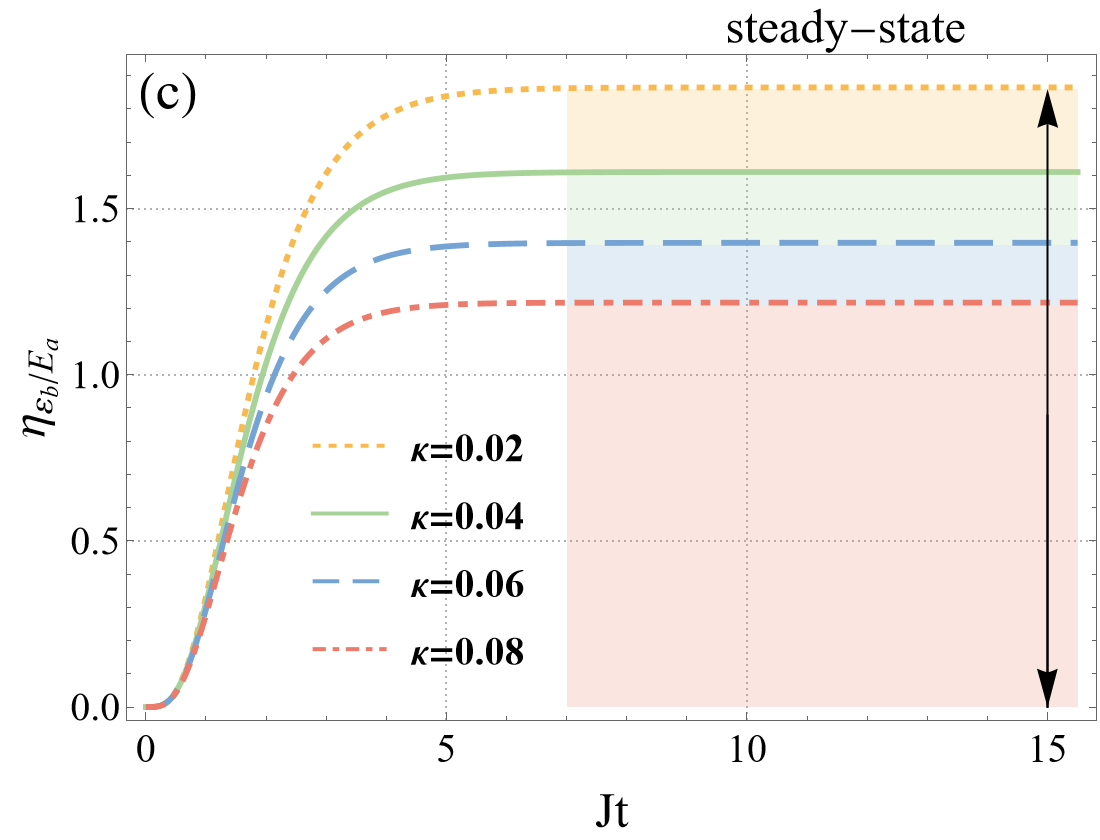}
	\includegraphics[width=0.825\columnwidth]{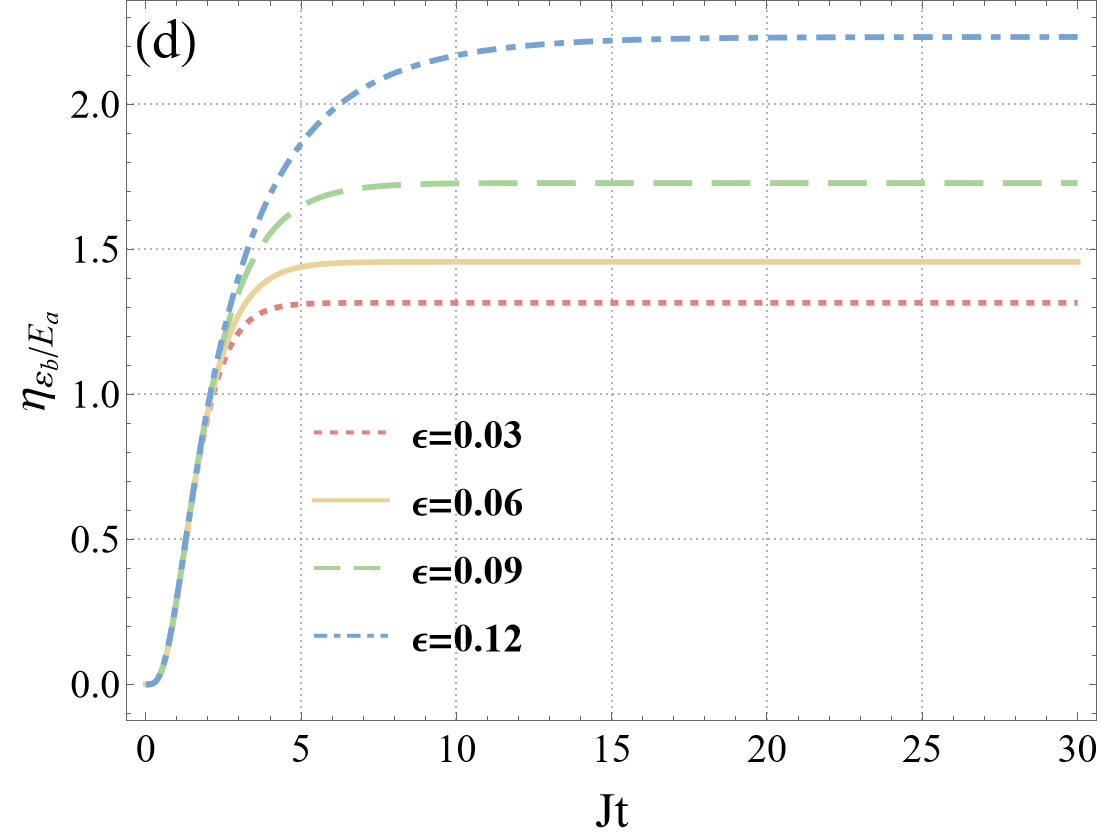}
	\caption{Figs (a) and (b) illustrate the functions of time $Jt$ evolution of the energy utilization ratio $\eta_{\varepsilon b/Eb}$ and the energy conversion ratio $\eta_{\varepsilon b/Ea}$ under various dissipation strengths $\kappa$. Figs (c) and (d) show how these ratios change with time under various driving strengths, with $\kappa$ held constant. In all panels, the horizontal axis denotes the rescaled time $Jt$. Other parameters are set as: $\kappa = \kappa_a = \kappa_b$, $\Gamma = \Gamma_a = \Gamma_b$, and $|J| = \Gamma/2 = 0.25\omega$.}
	\label{fig3}
\end{figure*}\par
{\color{black}Beyond the storage and ergotropy of QBs, their performance can be characterized by energy utilization ratio $\eta_{\varepsilon_b/E_b}$ and energy conversion ratio $\eta_{\varepsilon_b/E_a}$. 
%The former reflects the usable fraction of the total stored energy, indicating storage quality, while the latter evaluates the conversion ratio from input energy to available energy during charging, representing charging effectiveness. 
The first quantity, $\eta_{\varepsilon_b/E_b}$, measures the fraction of the total stored energy that can be extracted as useful work, thus reflecting the quality of energy storage. 
The second, $\eta_{\varepsilon_b/E_a}$, characterizes the efficiency of converting the input energy into extractable energy during the charging process, thereby quantifying the overall charging effectiveness. 
These quantities are defined by}
 \begin{equation}\label{eq18}
 	\eta_{\varepsilon _b/E_b}(t) =  
 	\frac{\varepsilon_b(t)}{E_b(t)},\\ \ \  \ \\ \eta_{\varepsilon _b/E_a}(t)=\frac{\varepsilon _b(t)}{E_a(t)}.
 \end{equation}
As shown in Figs.~\ref{fig3}(a) and~\ref{fig3}(c), we present the evolution of the QB energy utilization ratio $\eta_{\varepsilon_b/E_b}$ and the energy conversion ratio $\eta_{\varepsilon_b/E_a}$ with the characteristic time scale $Jt$ for different $\kappa$. 
%It is observed that both $\eta_{\varepsilon_b/E_b}$ and $\eta_{\varepsilon_b/E_a}$ exhibit a monotonic decrease as $\kappa$ increases. 
Both quantities exhibit a monotonic decrease with increasing $\kappa$, indicating that stronger dissipation suppresses the effective energy extraction and conversion efficiency of the QB. 
A closer inspection reveals distinct dynamical behaviors for the two ratios. $\eta_{\varepsilon_b/E_b}$ displays a non-monotonic dependence on $Jt$: it rises sharply to a peak value at early times, followed by a gradual decay toward a steady-state equilibrium. 
In contrast, the conversion ratio $\eta_{\varepsilon_b/E_a}$ increases monotonically with time, rapidly approaching a stable equilibrium value. 

%Further analysis reveals that $\eta_{\varepsilon_b/E_b}$ shows a non-monotonic evolution with the characteristic time scale $Jt$, initially rising sharply to a peak and then gradually decaying, eventually reaching a dynamic equilibrium. In contrast, $\eta_{\varepsilon_b/E_a}$ displays a monotonic increase, rising quickly at the initial stage and eventually stabilizing at a dynamic equilibrium.

Additionally, we also investigate the impact of driving field strength on the energy utilization ratio and energy conversion ratio of QB. As depicted in Figs.~\ref{fig3}(b) and~\ref{fig3}(d), both ratios $\eta_{\varepsilon_b/E_b}$ and $\eta_{\varepsilon_b/E_a}$ improve with increasing driving field strength $\epsilon$. Specifically, for higher values of $\epsilon$, $\eta_{\varepsilon_b/E_b}$ increases initially with $Jt$ before stabilizing, while for lower values of $\epsilon$, the ratio increases gradually, then decays slowly, and eventually reaches dynamic equilibrium. Further analysis indicates that the characteristic time $Jt$ required for the QB system to reach dynamic equilibrium lengthens with increasing $\epsilon$. This is due to the fact that a higher driving field enhances quantum fluctuations and the energy excitation level within the system, accelerating and improving the energy transfer from the charger to the battery. Furthermore, as illustrated in Figs.~\ref{fig3}(c) and~\ref{fig3}(d), both the energy utilization ratio $\eta_{\varepsilon_b/E_b}$ and the energy conversion ratio $\eta_{\varepsilon_b/E_a}$ exceed unity. This confirms that under nonreciprocal conditions, the QB system exhibits unidirectional energy transport without backflow during the charging process.\par

\subsection{Comparison between Single- and Two-Photon Nonlinear Driving in Nonreciprocal QB under Symmetric Dissipation}\label{Sec3B}\par
\begin{figure*}[htbp]
	\centering
	\includegraphics[width=0.85\columnwidth]{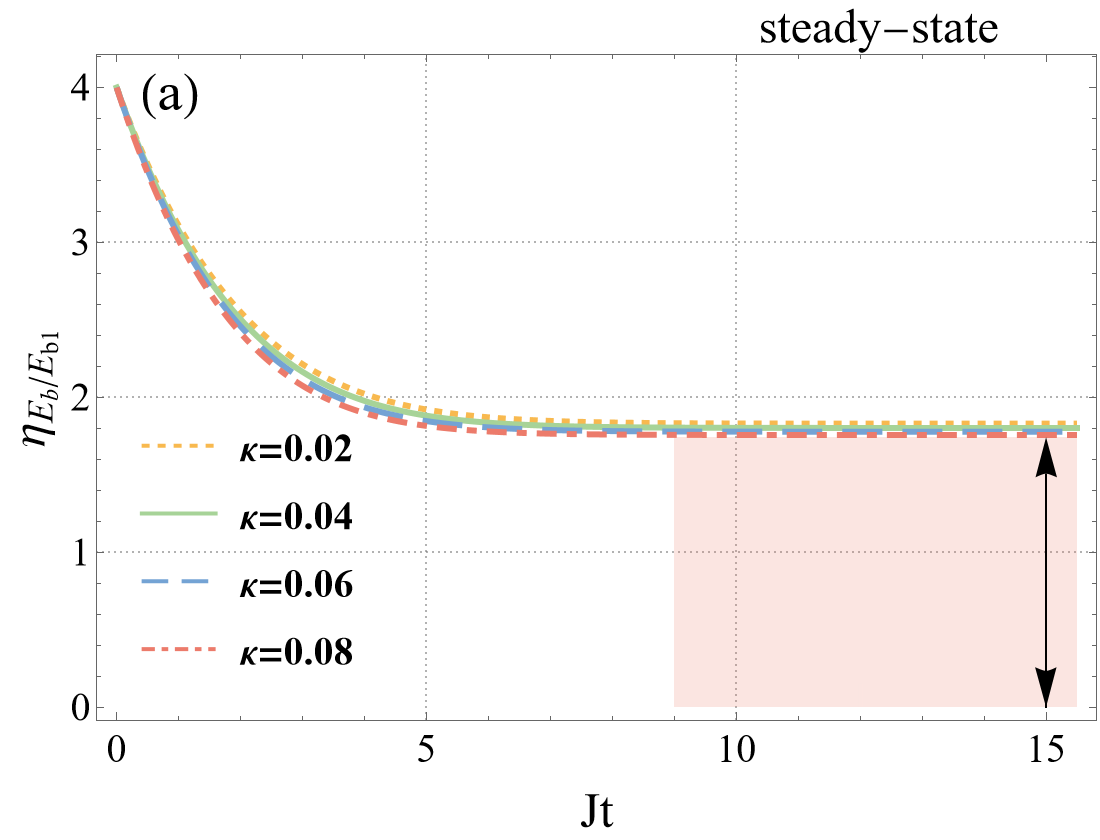}
	\includegraphics[width=0.825\columnwidth]{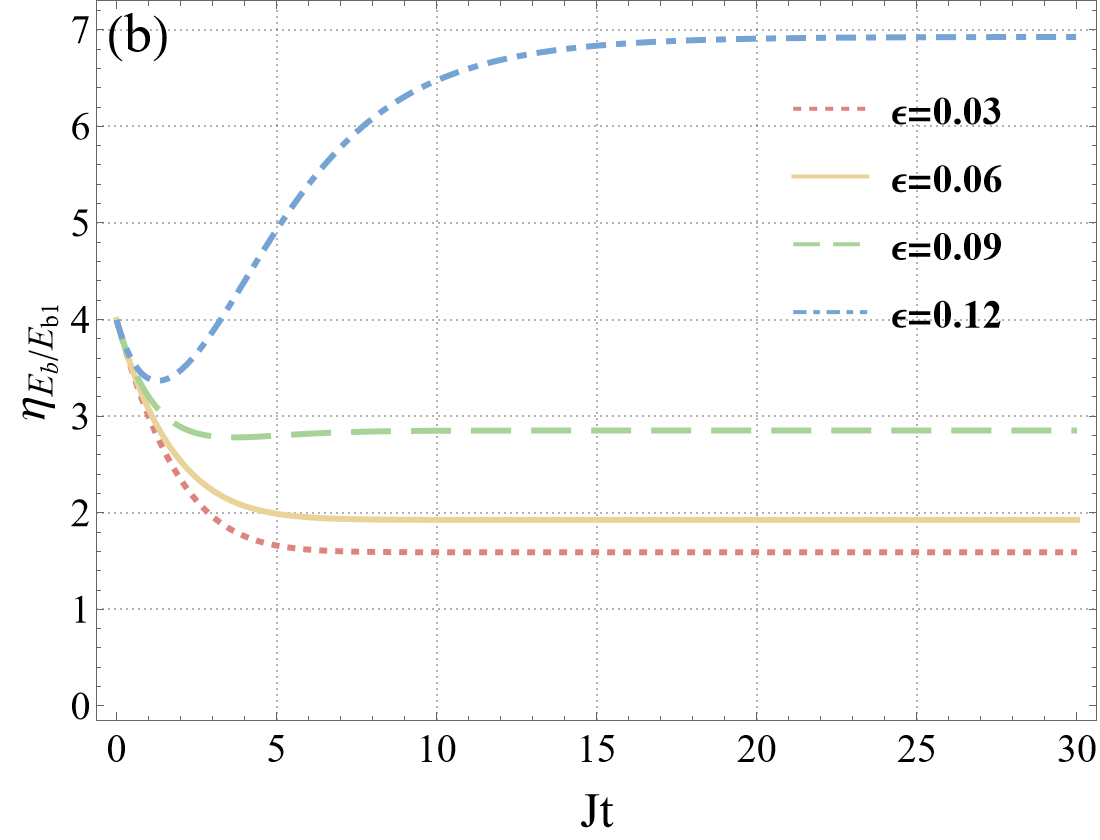}	\includegraphics[width=0.88\columnwidth]{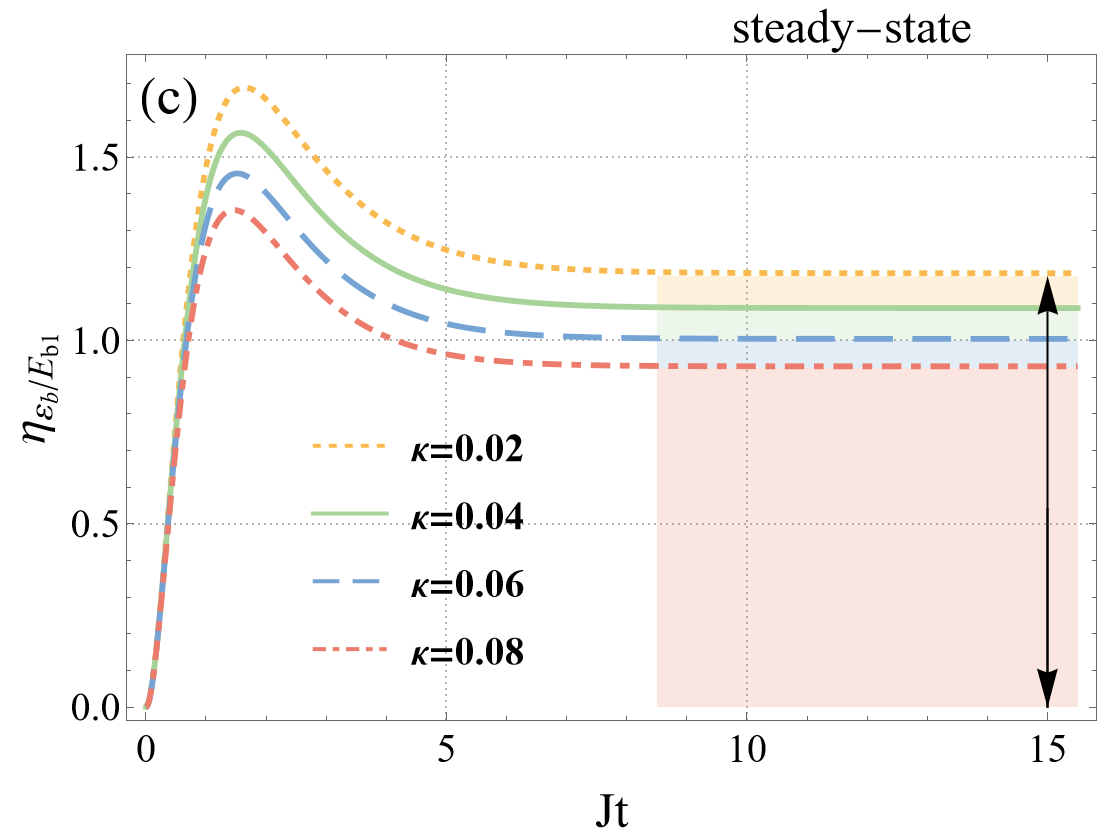}
	\includegraphics[width=0.825\columnwidth]{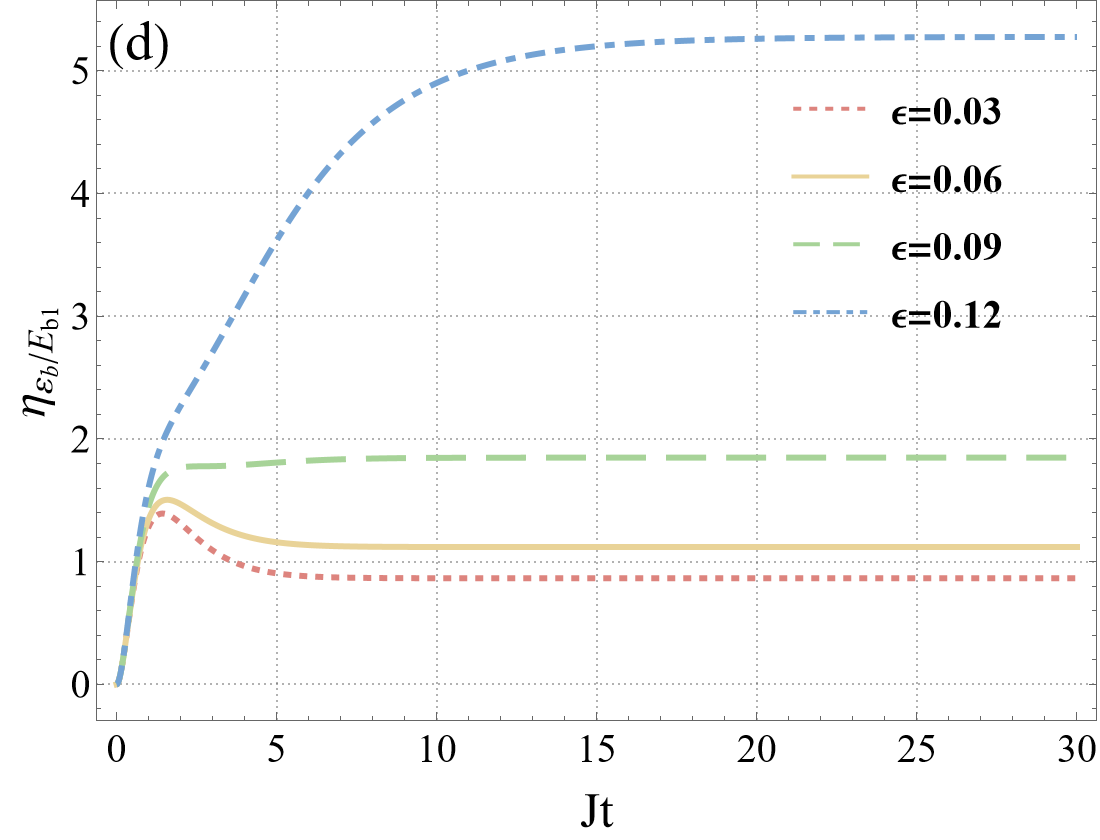}
	\caption{Figs (a) and (c) illustrate the dynamic evolution of the QB’s the relative storage ratio $\eta_{E_{b}/E_{b1}}$ and the relative ergotropy ratio $\eta_{\varepsilon_{b}/E_{b1}}$ under different $\kappa$ conditions, as a function of the characteristic scaling time $Jt$. Here, $\epsilon = 0.05 \omega$. Figures (b) and (d) explore the time evolution of these two quantities further, under a fixed dissipation strength of $\kappa = 0.06 \omega$, while varying the driving strength parameters $\epsilon/\omega = 0.03, 0.06, 0.09, 0.12$. Other parameters include: $\kappa = \kappa_{a} = \kappa_{b}$, $\Gamma = \Gamma_{a} = \Gamma_{b}$, $|J| = \Gamma /2$, and $|J| = 0.25 \omega$.}
	\label{fig4}
\end{figure*}\par

In recent years, the precise control of symmetries in nonreciprocal quantum systems has opened a new avenue for understanding and enhancing the energy-storage mechanisms of Quantum Batteries (QBs). Specifically, in nonreciprocal single-photon classically driven systems, the breaking of time-reversal symmetry enables highly efficient unidirectional energy transfer from the charger to the battery, leading to efficiency enhancements up to four times compared to conventional reciprocal architectures. Furthermore, QBs leveraging nonlinear two-photon {\color{black}driving} processes offer a distinct advantage, as the quadratic driving mechanism can significantly boost both energy storage capacity and ergotropy beyond the limits of linear coupling schemes.
Building on these foundational insights, we systematically evaluate the evolution of energy and ergotropy in nonreciprocal QBs realized through these two distinct driving mechanisms (single-photon vs. nonlinear two-photon). Our primary objective is to quantitatively compare their operational performance and identify the optimal regime for quantum energy transfer.
%In recent years, precise control of symmetries in nonreciprocal quantum systems has opened a new avenue for understanding the energy-storage mechanisms of QBs. In nonreciprocal single-photon classically driven systems, breaking time-reversal symmetry enables unidirectional energy transfer from the charger to the battery, thereby enhancing energy accumulation efficiency by up to four times compared to conventional systems. Furthermore, QBs based on nonlinear two-photon-driven processes can significantly boost energy storage and ergotropy through quadratic driving processes. Building on these insights, we systematically evaluate the evolution of energy and ergotropy in nonreciprocal QBs under these two distinct driving schemes and quantitatively compare their performance differences.

%Figs.\ref{fig4}~(a) and (c) illustrate the time-dependent dynamics of the relative storage ratio $\eta_{E_b/E_{b1}}$ and the relative ergotropy ratio $\eta_{\varepsilon_b/E_{b1}}$ under two distinct driving mechanisms, for different values of $\kappa$. 
Figs~\ref{fig4}(a) and~\ref{fig4}(c) display the time-dependent dynamics of the relative storage ratio $\eta_{E_b/E_{b1}}$ and the relative ergotropy ratio $\eta_{\varepsilon_b/E_{b1}}$ under two distinct driving mechanisms for various dissipation rates $\kappa$. 
%In the zero-temperature environment, the ergotropy of QBs under single-photon coupling is represented by $E_{b1}$\cite{kossakowski1972quantum}. 
In a zero-temperature environment, the ergotropy of the QB driven by single-photon is denoted as $E_{b1}$~\cite{kossakowski1972quantum}. 
Interestingly, the relative ratios exhibit opposing evolution behaviors in the two driving modes. Specifically, the relative storage ratio $\eta_{E_{b}/E_{b1}}$ monotonically decreases over time until reaching dynamic equilibrium, while the relative ergotropy ratio $\eta_{\varepsilon_{b}/E_{b1}}$ initially increases monotonically and then gradually decays, ultimately stabilizing at a steady state. The contrasting evolutionary behaviors stem from differences in the driving field and the corresponding environment. In the initial stage, the two-photon driving mechanism facilitates more efficient energy injection via nonlinear processes, resulting in a higher $E_b$ than $E_{b1}$. As time evolves, however, dissipation not only leads to energy leakage but also disrupts quantum squeezing, causing a rapid decay of $E_b$. Meanwhile, the evolution of $E_{b1}$ remains relatively stable. Consequently, the ratio $\eta_{E_b/E_{b1}}$ declines continuously until both systems attain dynamic equilibrium. On the other hand, due to the presence of quantum squeezing at the initial time, some energy is not extractable. As time progresses, dissipation-induced decoherence gradually transforms this portion of energy into extractable energy, which is reflected in the continuous increase of $\eta_{\varepsilon_{b}/E_{b1}}$ until it reaches dynamic equilibrium. As observed in Fig.~\ref{fig4}(c), when $\kappa \approx 0.61$, the ratio of the maximum extractable energy, $\eta_{\varepsilon_{b}/E_{b1}}$, is 1. 

Figs.~\ref{fig4}(b) and~\ref{fig4}(d) respectively illustrate the dynamic evolution of the relative storage ratio $\eta_{E_b/E_{b1}}$ and the relative ergotropy ratio $\eta_{\varepsilon_b/E_{b1}}$ over time under different $\epsilon$ conditions. From Fig.~\ref{fig4}(b), it is observed that, when the driving strength is sufficiently high, the relative storage ratio initially decreases and then gradually increases until it reaches equilibrium. Additionally, the higher the driving strength, the larger the relative storage ratio. Similarly, the relative ergotropy ratio of the battery increases as $\epsilon$ increases. Figs.~\ref{fig4}(c) and~\ref{fig4}(d) indicate that for dissipation rates $\kappa < 0.61$ or driving strengths $\epsilon > 0.03$, the nonlinear two-photon driving QB surpasses the single-photon QB in both relative storage and relative ergotropy.\par
\begin{figure*}[htbp]
	\centering
	\includegraphics[width=0.85\columnwidth]{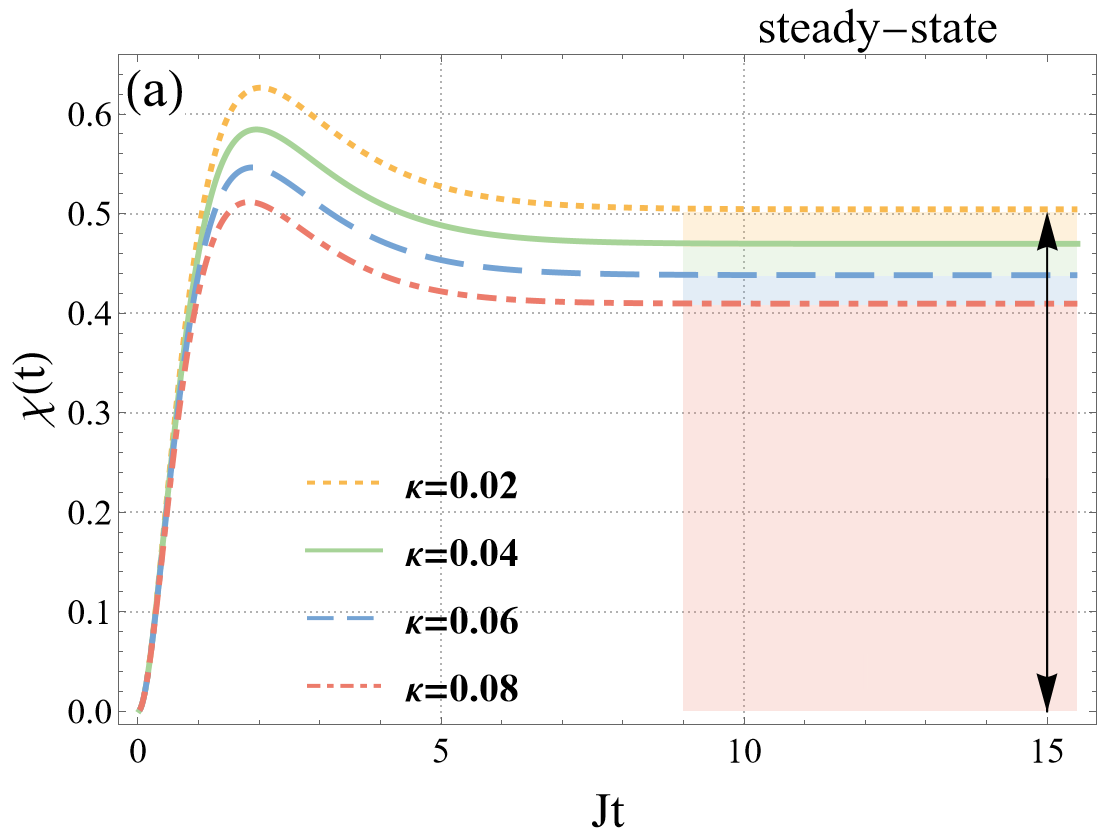}
	\includegraphics[width=0.825\columnwidth]{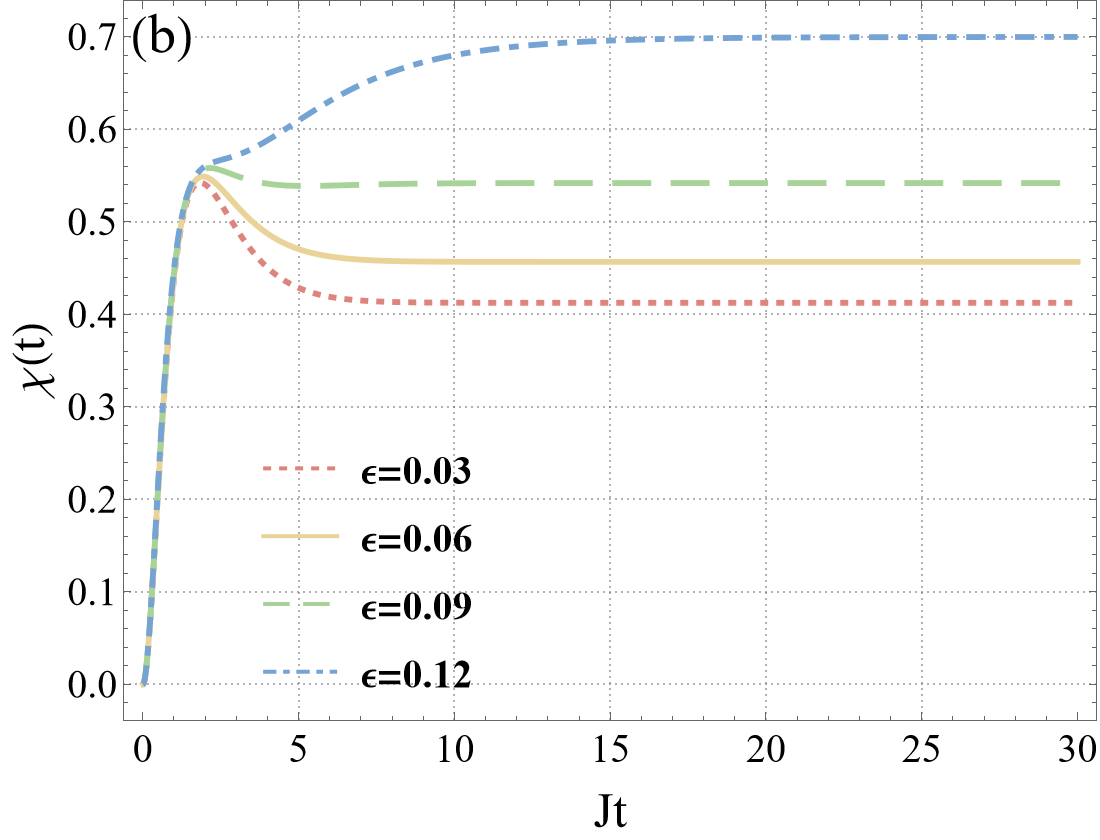}
	\caption{Figs (a) and (b) display the evolution of the efficiency ratio $\chi(t)$ between a nonlinear two-photon {\color{black}driving} QB and a single-photon QB under different conditions. Fig. (a) illustrates the relationship as a function of the dissipation rate $\kappa$, while Fig. (b) shows its dependence on the driving field strength $\epsilon$. The other parameters are set as follows: $\kappa = \kappa_a = \kappa_b$, $\Gamma = \Gamma_a = \Gamma_b$, $|J| = \Gamma / 2$, and $|J| = 0.25\omega$.}
	\label{fig5}
\end{figure*}\par
%From Figs.\ref{fig4}, it is demonstrated that the nonlinear two-photon driven QB outperforms the single-photon QB in both the relative energy ratio and the relative ergotropy ratio. 
From the results shown in Fig.~\ref{fig4}, it is evident that the nonlinear two-photon {\color{black}driving} QB exhibits superior performance over its single-photon counterpart in terms of both the relative energy and relative ergotropy ratios. 
To further quantify the performance differences, we define the relative ratio $\chi(t)$, which is given by:
\begin{equation}\label{eq19}
	\chi(t) =  
	\frac{\eta_{\varepsilon_{b}/E_{a}}}{\eta_{E_{b1}/E_{a1}}}.
\end{equation}
Surprisingly, as shown in Fig.~\ref{fig5}, $\chi(t)$ remains less than 1, indicating that the single-photon QB performs better than the nonlinear two-photon {\color{black}driving} QB. Furthermore, the evolution of $\chi(t)$ as a function of the characteristic time $Jt$ exhibits a two-stage behavior: an initial rapid increase followed by a gradual decay, ultimately stabilizing at a steady state. Specifically, under strong driving conditions, $\chi(t)$ rises sharply at first, with the rate of increase slowing down thereafter, while under weak driving, it shows a non-monotonic behavior, increasing initially before decaying. Regardless of the driving strength $\epsilon$ or dissipation rate $\kappa$, $\chi(t)$ remains consistently below 1.\par

\subsection{Nonreciprocal QB with Nonlinear Two-Photon Driving under Asymmetric Dissipation}\label{Sec3C}\par

%To investigate the impact of asymmetric dissipation on the performance of nonlinear two-photon-driven QB, we introduce local damping asymmetry based on the symmetric case discussed in Sec .~\ref {Sec3A}. Through the adjustment of the shared dissipative parameters $p_{a}$ and $p_{b}$, we realize a controllable asymmetry, and the ratio parameter is defined as follows:
To investigate the impact of asymmetric dissipation on the performance of nonlinear two-photon {\color{black}driving} QB, we introduce local damping asymmetry based on the symmetric case discussed in Sec .~\ref {Sec3A}. This is achieved by introducing a controllable asymmetry in the local damping rates. Specifically, the asymmetry is realized through the independent adjustment of the shared dissipative parameters $p_{c}$ and $p_{b}$ (corresponding to the charger and battery modes, respectively). 
%We define the dissipation asymmetry ratio $\mathcal{R}$ to quantitatively characterize this imbalance:
\begin{widetext}
Specifically, we introduce two dimensionless asymmetry parameters.
The intrinsic damping asymmetry ratio $\xi$, defined by the local decay rates of the two modes
\begin{equation}\label{eq20}
\xi  =\sqrt{\frac{\kappa _{a} }{\kappa _{b}} } .
\end{equation}
%We then redefine $p_{a}$ and $p_{b}$ through dimensionless parameters as:
The coupled dissipation ratio $x$, which governs the scaling of the shared dissipative parameters
\begin{equation}\label{eq21}
	\begin{aligned}
		p_{a} \to p_{a}\sqrt{x} ,\ \ \ \ \ \ p_{b} \to \frac{p_{b}}{\sqrt{x}} ,
	\end{aligned}
\end{equation}
with $\left | p_{a}  \right | =\left | p_{b}  \right | =1$, yielding the scaled damping rates $\Gamma _{a}\to \Gamma x  $ and $\Gamma _{b} \to \frac{\Gamma}{x} $. Under this framework, the analytical solutions for the energy storage $E_{b}$ and the ergotropy $\varepsilon_{b}$ of QB at the steady-state limit $t\to\infty$ are as follows: 
	\begin{subequations}\label{eq22}
		\begin{align}
			&E_{b}(t\to\infty ) =\frac{32\Gamma ^{2}\epsilon^{2}(\frac{\Gamma }{x}+2(x\Gamma +\kappa _{a})+\frac{\kappa _{a} }{\xi  ^{2} }  ) }{(x\Gamma +\kappa _{a} )((x\Gamma +\kappa _{a})^{2}-16\epsilon ^{2})((\frac{\Gamma }{x}+x\Gamma +\kappa _{a}+\frac{\kappa _{a} }{\xi ^{2} } )^{2}-16\epsilon ^{2})} ,\\
			&\varepsilon_{b}(t\to\infty ) =\frac{32\Gamma ^{2}\epsilon^{2}(\frac{\Gamma }{x}+2(x\Gamma +\kappa _{a})+\frac{\kappa _{a} }{\xi  ^{2} }  ) }{(x\Gamma +\kappa _{a} )((x\Gamma +\kappa _{a})^{2}-16\epsilon ^{2})((\frac{\Gamma }{x}+x\Gamma +\kappa _{a}+\frac{\kappa _{a} }{\xi ^{2} } )^{2}-16\epsilon ^{2})}.
		\end{align}
	\end{subequations}
\end{widetext}
Considering Eq~$\eqref{eq22}$ as functions of $x$, it is clear that these expressions are high-order rational functions of $x$. According to Galois theory, analytical solutions for extrema do not exist. To better understand the impact of asymmetry on the performance of QB, we systematically quantify the effects of the asymmetric parameters $x$ and $\xi$ on the energy storage $E_b$ and the ergotropy $\varepsilon_b$. \par
\begin{figure*}[htbp]
	\centering
	\includegraphics[width=1.9\columnwidth]{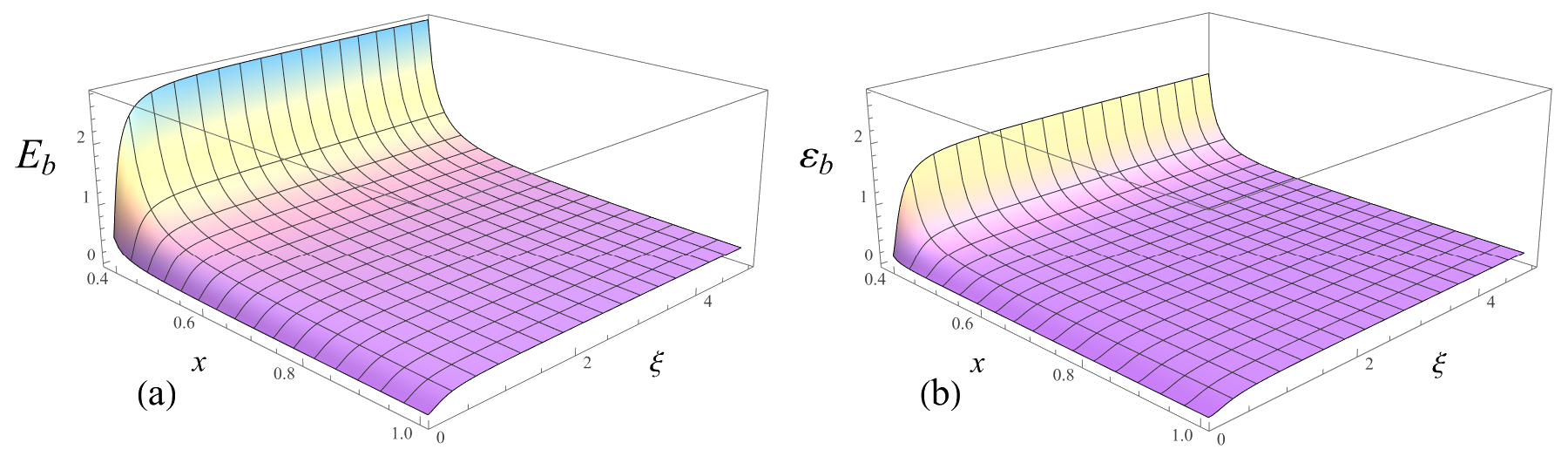}
	\includegraphics[width=0.1\columnwidth]{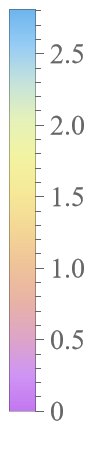}
	\caption{(a) and (b) illustrate the variation of the stored energy $E_b$ and the traversal ergotropy $\varepsilon_b$ of a nonlinear two-photon {\color{black}driving} QB with respect to the parameters $x$ and $\xi$, respectively. The remaining parameters are set as follows: $\kappa_a = 0.02\omega$, $\epsilon = 0.05\omega$, and $|J| = \Gamma/2 = 0.25\omega$.}
	\label{fig6}
\end{figure*}

To quantitatively examine the influence of dissipative asymmetry on the quantum battery (QB) performance, Fig.~\ref{fig6} displays the steady-state energy storage $E_b$ and the ergotropy $\varepsilon_b$ as functions of the asymmetry parameters $x$ and $\xi$. These results, obtained from the analytical expression in Eq.~\eqref{eq22}, clearly visualize the asymmetric response of the system under different dissipative configurations.

As shown in Fig.~\ref{fig6}, both $E_b$ and $\varepsilon_b$ decrease monotonically with increasing $x$. This behavior indicates that a stronger asymmetry between the charger and the battery (i.e., a larger difference between $\Gamma_a$ and $\Gamma_b$) enhances energy loss through the charger channel before efficient transfer to the battery can occur. Consequently, the unidirectional energy transport efficiency from the charger to the battery is reduced, leading to a degradation of the overall charging performance.

In contrast, for a fixed $x$, both $E_b$ and $\varepsilon_b$ increase gradually with $\xi$. Physically, a larger $\xi$ corresponds to a weaker coupling $\kappa_b$ between the battery and the shared dissipative reservoir, thereby suppressing direct energy leakage from the battery to the environment and improving the effective energy retention. Notably, the sensitivity of both $E_b$ and $\varepsilon_b$ to variations in $\xi$ is most pronounced when $x$ is small, i.e., when the intrinsic damping asymmetry between the charger and the battery is weak. In this near-symmetric regime, the decay rates $\Gamma_a$ and $\Gamma_b$ are comparable, and the system approaches a quasi-reversible coupling condition in which the symmetry of the shared reservoir dominates the global dissipation behavior. As a result, even slight modifications in $\xi$ can induce substantial changes in the charging performance.

In summary, the results reveal distinct roles of the two asymmetry sources: increasing the local damping asymmetry between the charger and the battery (\(x\)) generally deteriorates performance, whereas tuning the dissipative asymmetry associated with the shared environment (\(\xi\)) can be exploited to enhance energy storage efficiency. These insights provide useful guidance for engineering high-performance, symmetry-controlled quantum batteries(QBs).

\section{Conclusion}\label{Sec4}

In summary, we have proposed a nonlinear two-photon {\color{black}driving} nonreciprocal quantum battery (QB) that enables unidirectional energy transfer from the charger to the battery through engineered environmental dissipation. By solving the system's master equation, we derived the threshold condition for the existence of a steady-state solution, $\epsilon < \Lambda/4$, and systematically evaluated the QB performance under this constraint.

Our analysis reveals that, under symmetric and steady-state conditions, the local dissipation rate has a negligible influence on the relaxation time required to reach equilibrium. In contrast, increasing the driving field strength substantially prolongs the equilibration process while markedly enhancing the energy conversion and extraction efficiency. This indicates that strong driving fields play a crucial role in improving both the energy conversion and storage capabilities of the QB.

Compared with single-photon {\color{black}driving} counterparts, the nonlinear two-photon driving QBs exhibit superior energy storage capacity, particularly under strong driving conditions. Although the relative performance ratio remains below unity in certain dissipation and driving regimes, the nonlinear mechanism provides a clear advantage in terms of both energy storage and ergotropy. Moreover, in the presence of asymmetry, we demonstrate that optimizing the coupling between the QB and its dissipative environment can significantly boost the overall performance, highlighting the importance of symmetry control in nonreciprocal architectures.

Finally, the proposed QB scheme is experimentally accessible within several state-of-the-art quantum simulation platforms, including optoelectromechanical systems~\cite{barzanjeh2017mechanical}, superconducting quantum circuits~\cite{wang2016schrodinger}, and magnonic resonator systems~\cite{PhysRevLett.123.127202}. The abundant availability of nonreciprocal elements and nonlinear resonators in these platforms provides a promising route toward experimental realization and verification of the present results.

%========================================================================================   

\section*{ACKNOWLEDGMENTS}\par
C.R. was supported by the National Natural Science Foundation of China (Grants No. 12075245, 12421005, and No. 12247105), Hunan provincial major sci-tech program (No. 2023ZJ1010), the Natural Science Foundation of Hunan Province (2021JJ10033), the Foundation Xiangjiang Laboratory (XJ2302001), and Xiaoxiang Scholars Program of Hunan Normal University.\par

\appendix

\numberwithin{equation}{section}
\renewcommand\theequation{\Alph{section}.\arabic{equation}}
\numberwithin{equation}{section}
\renewcommand\theequation{\Alph{section}.\arabic{equation}}
\section{Steady-state solutions of quantum battery system}\par
From the master equation, the steady-state condition $\dot{\tilde{\hat{\rho}}} = 0$ yields the following solutions for the correlation functions:
\begin{widetext}
\begin{equation}
	\begin{aligned}
		&\left \langle \hat{a}^{\dagger }\hat{a}  \right \rangle_{\infty }  =\frac{8\epsilon ^{2} }{\Lambda ^{2}-16\epsilon ^{2}} , \ \ \ \ \left \langle \hat{b}^{\dagger} \hat{b}  \right \rangle_{\infty }=\frac{32 \Gamma^2 \epsilon^2 (\Delta + 2 \Lambda) \mu \mu^*}{\Delta (\Delta - 4 \epsilon + \Lambda) (\Delta + 4 \epsilon + \Lambda) (\Lambda^2-16 \epsilon^2 )},
		\\
		&\left \langle \hat{a}^{\dagger }\hat{a}^{\dagger }\right \rangle_{\infty }=\left \langle \hat{a}\hat{a}\right \rangle^{*}_{\infty } = \frac{2 i\mathrm{e}^{i\theta }\epsilon \Lambda  }{ (\Lambda^2-16 \epsilon^2 )}, \ \ \ \ \left \langle \hat{b}^{\dagger} \hat{b}^{\dagger}  \right \rangle_{\infty } =\left \langle \hat{b}\hat{b}\right \rangle^{*}_{\infty }= - \frac{8 i e^{-i \theta} \Gamma^2 \epsilon \left(16 \epsilon^2 + \Lambda \left( \Delta + \Lambda \right) \right) \mu^2}{\Delta (\Delta - 4 \epsilon + \Lambda) (\Delta + 4 \epsilon + \Lambda) (16 \epsilon^2 - \Lambda^2)},\\ 
		&\left \langle \hat{a}^{\dagger }\hat{b}  \right \rangle_{\infty }=\left \langle \hat{a}\hat{b}^{\dagger }  \right \rangle^{*}_{\infty }=\frac{16 \Gamma \epsilon^2 (\Delta + 2 \Lambda)\mu^*}{(16 \epsilon^2 - \Lambda^2) ( 16 \epsilon^2-(\Delta+\Lambda)^2)} , \ \ \ \ \left \langle \hat{a}^{\dagger }\hat{b}^{\dagger }  \right \rangle_{\infty }= \left \langle \hat{a}\hat{b}\right \rangle^{*}_{\infty } = \frac{4 i e^{-i \theta} \Gamma \epsilon \left(16 \epsilon^2 + \Delta \Lambda + \Lambda^2 \right) \mu}{\left(16 \epsilon^2 - \Lambda^2\right) \left( 16 \epsilon^2-(\Delta+\Lambda)^2 \right)}.\\
	\end{aligned}
\end{equation}
\end{widetext}
In the symmetric case $\Delta=\Lambda$, and using Eqs.~$\eqref{eq13}$ and $\eqref{eq15}$, the corresponding steady-state battery energy and ergotropy become:
\begin{widetext}
	\begin{equation}
		\begin{aligned}
	E_{b}(\infty)/\omega&=\frac{24\Gamma ^{2}\epsilon^{2}}{64\epsilon ^{4}-20\epsilon ^{2}\Lambda^{2}+\Lambda^{4}},\\
	\varepsilon(\infty)/\omega&=\frac{24\Gamma ^{2}\epsilon^{2}}{64\epsilon ^{4}-20\epsilon ^{2}\Lambda^{2}+\Lambda^{4}}-\frac{1}{2}\left(-1+\sqrt{\frac{-64 \Gamma^{4}\epsilon ^{2}+96 \Gamma^{2}\epsilon ^{2}\Lambda^{2}+64 \epsilon ^{4}\Lambda^{2}-2 \epsilon^{2}\Lambda^{4}+\Lambda^{6}}{64 \epsilon^{4}\Lambda^{2}-2 \epsilon^{2}\Lambda^{4}+\Lambda^{6}}}\right).\\
\end{aligned}
\end{equation}
\end{widetext}

\bibliographystyle{apsrev4-1}
\bibliography{ref}

\end{document}